\documentclass[letterpaper,twocolumn,10pt]{article}
\usepackage{usenix-2020-09}
\usepackage{paperspecific}

\usepackage{tikz}
\usepackage{amsmath}

\usepackage{filecontents}

\begin{document}
%-------------------------------------------------------------------------------

\begin{acronym}
	\acro{csr}[CSR]{Control and Status Register}
	\acro{encid}[EncID]{Enclave Identifier}
	\acro{epcm}[EPCM]{Enclave Page Cache Map}
	\acro{isa}[ISA]{Instruction Set Architecture}
	\acro{loc}[LoC]{Lines of Code}
	\acro{mee}[MEE]{Memory Encryption Engine}
	\acro{mmu}[MMU]{Memory Management Unit}
	\acro{mpu}[MPU]{Memory Protection Unit}
	\acro{pmp}[PMP]{Physical Memory Protection}
	\acro{os}[OS]{operating system}
	\acroplural{os}[OSes]{operating systems}
	\acro{pmh}[PMH]{Page Miss Handler}
	\acro{pte}[PTE]{\pt entry}
	\acroplural{loc}[LoC]{Lines of Code}
	\acroplural{pte}[PTEs]{\pt entries}
	\acro{rvas}[RVAS]{RISC-V Authenticryption Shield}
	\acro{rt}[RT]{runtime}
	\acro{rtid}[RTID]{Runtime Identifier}
	\acro{sev}[SEV]{Secure Encrypted Virtualization}
	\acro{sid}[SID]{Session Identifier}
	\acro{sm}[SM]{Security Monitor}
	\acro{tcb}[TCB]{Trusted Computing Base}
	\acro{tlb}[TLB]{Translation Lookaside Buffer}
	\acro{tee}[TEE]{Trusted Execution Environment}
	\acro{tbl}[TBL]{Tweak Buffer List}
	\acro{tc}[TC]{Tweak Cache}
	\acro{llc}[LLC]{Last Level Cache}
	\acro{soc}[SoC]{System on Chip}
\end{acronym}

%don't want date printed
\date{}

\title{\Large \bf SERVAS! Secure Enclaves via \acl{rvas}}

\ifanonymous
\else
\author{
{\rm Stefan Steinegger}\\
Graz University of Technology
\and
{\rm David Schrammel}\\
Graz University of Technology
\and
{\rm Samuel Weiser}\\
Graz University of Technology
\and
{\rm Pascal Nasahl}\\
Graz University of Technology
\and
{\rm Stefan Mangard}\\
Graz University of Technology\\
Lamarr Security Research
\\
}

\fi % ifanonymous

\maketitle

%-------------------------------------------------------------------------------
\begin{abstract}
%-------------------------------------------------------------------------------
Isolation is a long-standing challenge of software security. 
Traditional privilege rings and virtual memory are more and more augmented with concepts such as capabilities, protection keys, and powerful enclaves.
At the same time, we are evidencing an increased need for physical protection, shifting towards full memory encryption schemes.
This results in a complex interplay of various security mechanisms, increasing the burden for system architects and security analysts.

In this work, we tackle the isolation challenge with a new isolation primitive called authenticryption shield that unifies both traditional and advanced isolation policies while offering the potential for future extensibility.
At the core, we build upon an authenticated memory encryption scheme that gives cryptographic isolation guarantees and, thus, streamlines the security reasoning.
We showcase the versatility of our approach by designing and prototyping \ourEnclave{} -- an innovative enclave architecture for \riscv. 
Unlike current enclave systems, \ourEnclave facilitates efficient and secure enclave memory sharing.
While the memory encryption constitutes the main overhead, 
entering or exiting a \ourEnclave enclave requires only \benchriscventerexitgetpid{} of a simple syscall, instead of 71x for Intel SGX.

\end{abstract}

\section{Introduction}
Modern IT systems need to secure a multitude of applications against software attacks.
Unfortunately, software vulnerabilities penetrate the whole software stack, affecting not only application software~\cite{DBLP:conf/sp/SzekeresPWS13,DBLP:conf/ccs/BletschJFL11,DBLP:conf/ccs/CheckowayDDSSW10,DBLP:conf/ccs/Shacham07} but also the \ac{os} itself~\cite{DBLP:conf/apsys/ChenMWZZK11,stagefright2015,androidzeroday2019,ioszeroday2020,ioszerodayradio2020}.
In the worst case, attackers can gain full control over the device~\cite{androidzeroday2019,ioszeroday2020,ioszerodayradio2020}.

To reduce the impact of vulnerability exploitation, different isolation mechanisms are deployed on all levels, \eg to separate privileges~\cite{DBLP:conf/asplos/DautenhahnKDCA15}, isolate individual processes~\cite{DBLP:conf/bigcom/ZhuHX18}, 
protect virtual machines~\cite{AMDSev,AMDMemoryEncryption,AMDSevEs,AMDSevSnp},
and segregate applications into smaller parts, also denoted as in-process isolation.
Typical in-process isolation mechanisms are segmentation~\cite{IntelSDM} and capabilities~\cite{DBLP:conf/isca/WoodruffWCMADLNNR14}, memory coloring using, \eg protection keys~\cite{%
DBLP:conf/uss/Vahldiek-Oberwagner19,% % ERIM
DBLP:conf/usenix/HedayatiGJCSSM19,%   % Hodor
DBLP:conf/uss/SchrammelWSS0MG20}, %    % Donky
or enclaves.
Enclaves give strong security guarantees even in the event of a system compromise and found ample resonance both in academia~\cite{%
DBLP:conf/ics/SuhCGDD03,%              % AEGIS
IBMSecureBlue,%                        % IBM SecureBlue++
DBLP:conf/micro/EvtyushkinEOPGR14,%    % ISO-X
DBLP:conf/uss/CostanLD16,%             % Sanctum
DBLP:conf/isca/PetersLVPK18,%          % Bastion
DBLP:conf/asplos/TaassoriSB18,%        % VAULT
DBLP:conf/micro/BourgeatLWZAD19,%      % MI6
PenglaiEnclaves,%                      % Penglai Enclaves
DBLP:conf/ndss/EldefrawyTFP12,%        % SMART
DBLP:conf/acsac/GotzfriedMCMFV15,%     % Soteria
DBLP:conf/eurosys/KoeberlSSV14,%       % TrustLite
DBLP:conf/dac/BrasserMSWK15,%          % TyTAN
DBLP:journals/tissec/NoormanBMPMPVGM17,% Sancus 2.0
DBLP:conf/ndss/WeiserWBMMS19,%         % Timber-V
DBLP:conf/eurosys/LeeKSAS20,%          % Keystone
bahmani2020cure,%                      % CURE
DBLP:journals/corr/abs-2010-08440}     % Elasticlave
and industry~\cite{%
ARMTrustZone,%
ARMTrustZone_m,%
DBLP:conf/isca/McKeenABRSSS13,%        % SGX
SiFiveMultizone}.%                     % MultiZone

In addition, cloud computing scenarios increasingly demand physical protection, for which transparent memory encryption is being deployed on modern CPUs~\cite{DBLP:conf/isca/McKeenABRSSS13,AMDSevSnp}.
While memory encryption works well for small workloads, worst-case throughput penalties surpass 400\% for Intel SGX~\cite{DBLP:conf/srds/GottelPRVFPS18}.

Unfortunately, reasoning about the security of the interplay between these mechanisms is becoming more and more complex when considering the whole zoo of isolation mechanisms. 
For example, the security of an application might depend on protection keys in combination with the \acf{mmu} and the memory mappings configured by the operating system~\cite{kolanski2011verification,DBLP:conf/tase/JomaaNGH16,DBLP:conf/bigcom/ZhuHX18}.
Unifying these isolation mechanisms is desirable from a security standpoint.
On the other hand, most isolation mechanisms cover only a subset of isolation challenges.
For example, SGX enclaves can isolate small portions of unprivileged user code, but their underlying memory encryption is not utilizable for other purposes.

\textit{In this work}, we pursue a three-fold goal: 
First, we simplify the overall security reasoning by introducing a strong and generic isolation primitive.
Second, we explore the synergies and features our isolation primitive offers over traditional isolation mechanisms.
Third, we design a secure and feature-rich enclave architecture -- arguably the most intricate endeavor.

\subsubsection{New Isolation Primitive.}
We introduce a new isolation primitive that unifies various isolation policies and is denoted as \acf{rvas}.
At its core, \ac{rvas} uses memory encryption to map isolation properties to the well-studied field of cryptography, counteracting both physical and software attacks.
More specifically, if decryption succeeds, we know that the CPU and the memory are in a particular state.
\textit{Thus, \ac{rvas} achieves memory isolation with cryptographic guarantees.}

We design \ac{rvas} as an extension to the \riscv instruction set architecture~(ISA).
\ac{rvas} builds upon an authenticated memory encryption scheme whose associated data input, which we call encryption tweak, is exposed to software.
By controlling the encryption tweak, one can achieve domain separation and, thus, enforce a variety of different isolation mechanisms simultaneously, \eg privilege separation, process isolation and virtual memory protection, segmentation, and page coloring.
Traditionally, each of these mechanisms requires to securely store and manage trusted metadata (\eg the address mapping or the page colors).
\ac{rvas} implicitly secures this metadata by feeding it into the encryption tweak.

Achieving a proper generalization of encryption tweaks is non-trivial, however.
Hence, \rvas composes the tweak of both critical CPU-internal data and software-defined values to account for a broad range of use cases.

\subsubsection{\ourEnclave Enclaves.}
To showcase the versatility of \rvas, we use it to design a novel enclave system dubbed \ourEnclave, which has several advantages over existing enclave systems:

\ourEnclave protects enclaves against software and physical attacks by relying on a single mechanism, namely RVAS' cryptographic isolation.
In contrast, Intel SGX protects enclaves against \textit{physical} attacks via memory encryption while also preventing \textit{software} attacks through a trusted metadata storage (\ie the so-called EPCM~\cite{IntelSDM}).
The EPCM needs to be maintained for each enclave page and kept in sync with the \ac{tlb}, for which SGX requires \ac{tlb} flushes upon enclave invocations~\cite{DBLP:journals/iacr/CostanD16}.

Our \rvas design makes the EPCM trusted metadata storage obsolete, which yields three advantages: First, we avoid \ac{tlb} flushes and thus, achieve better enclave invocation performance.
Second, we remove trust from the entire address translation, \ie the \ac{mmu} and the \ac{tlb} configuration,
such that our security reasoning boils down to a proper encryption tweak management.
Third, Intel SGX fixes the amount of encrypted enclave memory at boot time to typically \SI{128}{\mega\byte}~\cite{DBLP:conf/closer/GjerdrumPJJ17}.
\ac{rvas} encryption, in contrast, can be applied to the whole DRAM and also to non-enclave code.

\ourEnclave introduces the novel concept of secure sharing of enclave memory.
Secure sharing is a key requirement for many application scenarios, but it is impractical with current enclave systems (e.g., Intel SGX).
Thus, enclave-to-enclave communication currently demands costly encryption and copying operations in software.
\ourEnclave enables zero-cost secure data exchange by sharing specific encryption tweaks between eligible enclaves.
Moreover, selective code deduplication reduces memory demands if the same enclave is invoked multiple times.

\ourEnclave is compatible with advanced enclave features such as dynamic enclave memory, swapping, and multithreading.
As a proof of concept, we prototype \ac{rvas} on an FPGA by using the \ariane \riscv CPU together with an openly available encryption core.
A small stateless \acf{sm} running in \riscv machine mode\footnote{This is loosely comparable to CPU microcode used for Intel SGX logic.} ensures a proper tweak configuration for strongly-protected \ourEnclave enclaves.
We show that entering and exiting \ourEnclave enclaves only takes \benchriscventerexitgetpid{} the time of a syscall.
Our evaluation indicates an overhead between \SI{\macrooverheadmin}{\percent} and \SI{\macrooverheadmax}{\percent} over the used encryption core for a broad selection of macrobenchmarks.
We plan to open-source our prototype.

In summary, our contributions are as follows:
\begin{compactitem}
	\item A generic isolation primitive using authenticated memory encryption denoted as \acl{rvas}.
	\item A novel enclave architecture called \ourEnclave that leverages the \acl{rvas}.
	\item A novel and fast and secure memory sharing mechanism between enclaves.
	\item An evaluation of \ourEnclave in a set of micro- and macrobenchmarks.
\end{compactitem}

\subsubsection{Outline.}
The remaining paper is structured as follows:
In \Cref{sec:security_properties}, we discuss challenges of memory isolation.
In \Cref{sec:rvas}, we discuss our generic isolation primitive \ac{rvas}.
In \Cref{sec:enclaveDesign}, we present our \ourEnclave design, discuss how we use our components to build enclaves, and elaborate key points of their life-cycle.
\Cref{sec:implementationDetails} gives implementation details of \ac{rvas} and \ourEnclave. 
It elaborates on the specifics of the instruction set extension, the construction of the tweak, and the API of our \acl{sm}.
Additionally, this section discusses how tweaks are cached and the benefits of separating them into a tweak cache.
In \Cref{sec:securityanalysis}, we give a security analysis.
We evaluate our prototypes of \rvas and \ourEnclave in \Cref{sec:benchmarks}.
We discuss related work in \Cref{sec:related_work}, future work in \Cref{sec:additionalFeatures}, and conclude the paper in \Cref{sec:conclusion}.

\section{Challenges of Memory Isolation}\label{sec:security_properties}
The fragmentation of isolation mechanisms makes it hard to analyze their security, especially if multiple mechanisms are combined.
In this section, we give an overview of the most widely used isolation schemes and present their challenges concerning security and functional limitations we want to overcome.
This overview paves the way for understanding how the \ac{rvas} design can solve these challenges in general (\cf \Cref{sec:rvas}) and for enclaves in particular (\cf\Cref{sec:enclaveDesign}).

\subsection{Process Isolation}
Process isolation comprises separating privileges between an operating system and user processes and isolating processes from each other.
Privilege separation is achieved via privilege rings -- one of the coarsest protection mechanisms available in CPUs that protect privileged CPU resources from unprivileged access.
However, to be secure, privileged software also needs to protect its own memory and user memory using the virtual memory subsystem, as follows: 

\textbf{Challenge \ChalProcIso:} \textit{``The privileged software must ensure that the virtual memory mappings of \textbf{all} unprivileged processes (i) cannot access privileged memory, and (ii) are not unintentionally aliasing with each other.''}

Unfortunately, analyzing the security of process isolation is far from trivial~\cite{kolanski2011verification,DBLP:conf/tase/JomaaNGH16,DBLP:conf/bigcom/ZhuHX18} and requires a deep understanding of the memory management facilities of modern operating systems.
Moreover, the virtual memory mapping is a favored attack target~\cite{DBLP:conf/ndss/DaviGLS17} since a single bit flip in the page tables can suffice for privilege escalation attacks~\cite{DBLP:conf/ccs/VeenFLGMVBRG16}.

\subsection{In-Process Isolation}
\subsubsection{Segmentation} is an isolation mechanism to more finely separate parts within an application from each other. 
It usually confines memory accesses within predefined address ranges. 
Segmentation also forms the basis of hardware capabilities~\cite{DBLP:conf/isca/WoodruffWCMADLNNR14}.
However, these systems are not suitable for enforcing policies across application boundaries.
\Eg a segment to protect an application's cryptographic key is typically not respected by other applications (in case of shared memory) or by the OS. 

\textbf{Challenge \ChalSegmentMeta:} \textit{``Segmentation should also allow flexible cross-application policies.''}

\paragraph{Memory Coloring} is another in-process isolation mechanism labeling each memory block with a different ``color''.
The memory only becomes accessible to the application if the corresponding color is loaded in a special register.
Unfortunately, the number of colors is often quite limited~\cite{DBLP:conf/usenix/ParkLXMK19}, inhibiting fine-grain use cases.
Moreover, memory coloring is not enforced across application boundaries, making it unsuitable for sharing data with other applications.

\textbf{Challenge \ChalMemoryColor:} \textit{``Memory coloring should provide significantly more colors and also allow cross-application policies.''}

\subsection{Enclaves}\label{sec:enclave_challenges}
\subsubsection{Memory Mapping.}
Enclaves present an inverse problem of process isolation: an unprivileged software -- the enclave -- wants to protect itself from privileged software that manages the enclave's virtual memory mapping. This leads to:

\textbf{Challenge \ChalAddressTranslation:} \textit{``The memory mapping of enclaves must be protected against privileged software.''}

Protecting memory mappings against privileged attackers is challenging because privileged software is in legitimate charge of managing memory.
For example, if the \ac{os} maps memory to a wrong location, the enclave could be tricked into accessing the wrong buffer, leaking secrets, or corrupting its sensitive memory.
Also, by mapping code pages in a bogus way, the enclave could be forced to execute unintended functions or bypass security checks. 
Furthermore, manipulation of page table attributes could violate security assumptions and, \eg cause data pages to become executable.
Three security invariants need to hold, to prevent such attacks:

\textbf{Attribute Invariant} \IA: \textit{``Enclave pages must only be mapped with their intended page table attributes.''}

\textbf{Spatial Invariant} \IS: \textit{``A physical enclave page must only be mapped to its corresponding virtual page.''}

\textbf{Temporal Invariant} \IT: \textit{``At any time for every virtual enclave page, there must be at most one valid physical page mapping.''}

The \textit{temporal invariant} specifically addresses double mapping attacks:
If an attacker obtains two valid mappings for the same virtual enclave page, the \ac{os} could silently replace the underlying physical page in order to replay old data to the enclave and, thus, tamper with its execution.
This is especially relevant for dynamic enclave memory, swapping, and shared memory, where the memory utilization varies over time.

\paragraph{Protected Sharing}
is an important feature required for the interaction between applications. 
In an unprotected setting, this interaction is achieved via shared memory. 
However, in the context of enclaves, there is a hard isolation boundary that prohibits secure, shared memory by design.
In Intel SGX, data exchange is only possible via untrusted application memory, and enclaves need to manually encrypt data being sent.

\textbf{Challenge \ChalProtectedSharing:} \textit{``Shared memory must allow for efficient and confidential interaction between different enclaves.''}

\subsection{Memory Encryption} Apart from these architectural challenges, the DRAM presents a substantial physical attack surface, allowing passive~\cite{DBLP:conf/IEEEares/LindenlaufHS15,busprobing,DBLP:conf/uss/HaldermanSHCPCFAF08} and active~\cite{DBLP:conf/isca/KimDKFLLWLM14} attacks to infer or tamper with secret data stored in memory.
To isolate data from physical attacks and ensure its integrity, encrypting and authenticating the DRAM is necessary.
Memory encryption should not be restricted to specific code (\eg enclaves) or specific parts of the DRAM.

\textbf{Challenge \ChalMemoryEncryption:} \textit{``The DRAM shall be hardened against active and passive physical attacks.''}

\section{\acf{rvas}}\label{sec:rvas}
\rvas harnesses authenticated memory encryption as a single, generic mechanism to cryptographically enforce the challenges for memory isolation expressed in \Cref{sec:security_properties}.
At its core, we use a \ac{mee} for encrypting the DRAM and incorporate a security context into its tweak input (\ie the associated data).
If encrypted data is accessed with the wrong security context, the \ac{mee} triggers an authentication error.
Since the \ac{mee} gives cryptographic security guarantees for detecting authentication issues, the security argumentation for all the isolation policies boils down to one question: \textit{Who controls the security context?}

The composition of the security context arguably lies at the heart of \rvas.
For readability, we also call it \enquote{tweak} in the rest of the paper.
The tweak consists of both software- and CPU-defined components, allowing for fine-grained, unforgeable isolation.
We designed an \acf{isa} extension for configuring the tweak (\cf \Cref{sec:implementationDetails}), which makes \rvas a powerful and generic isolation primitive based on strong cryptographic segregation.

In this section, we first discuss the composition of the \rvas tweak.
Second, we sketch how \rvas solves the challenges defined in \Cref{sec:security_properties}.
We will concretely instantiate and analyze \ourEnclave enclaves addressing challenges \ChalSegmentMeta -- \ChalMemoryEncryption in \Cref{sec:enclaveDesign,sec:securityanalysis}.
Finally, we highlight requirements for the \ac{mee} to support \rvas. 

\subsection{\ac{rvas} tweak design}\label{sec:rvas:rvasTweakDesign}
A core contribution of \rvas lies in the way we compose the tweak used by the \acf{mee} from both software-defined values and CPU-defined security state.

\begin{figure}
	\includegraphics[width=0.8\columnwidth]{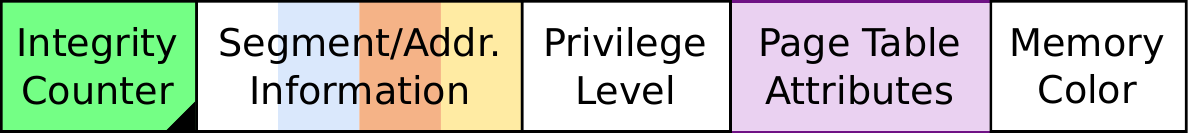}
	\vspace{-0.2em}
	\centering
	\captionof{figure}{Overview of the \ac{rvas} tweak composition.}
	\label{fig:tweakoverview}
\end{figure}

Our tweak design is depicted in \Cref{fig:tweakoverview} and comprises hardware-managed integrity counters, segment and address information, privilege levels, page table attributes, as well as software-defined memory colors.
All tweak components can be selectively enabled, depending on the specific use case (\eg unprotected applications, privilege separation only, enclaves).

\subsubsection{Integrity Counter.} 
The \ac{mee} maintains integrity counters for each memory block, which it increments at each write operation.
Integrating the counter into the tweak ensures that the correct memory block is used at any time.
This counter prevents replay attacks where an attacker with access to the physical memory reverts a memory block to its former state.

\subsubsection{Segmentation \& Address Information.}
This part of the tweak holds metadata about the address being accessed and whether it matches software-defined segments that can be configured at each privilege level. 
The address information can hold an absolute address or an address offset relative to one of the segments.
The segment information is represented as a segment bitmap indicating whether an address lies within one or multiple segments. 

\subsubsection{Privilege Level.}
This part of the tweak holds the current CPU privilege level (\eg M-mode, S-mode, U-mode) and ensures that memory is only accessible at a specific level.

\subsubsection{Page Table Attributes}
cover read, write and execute permissions, amongst others.
The inclusion of the page table attributes in the tweak ensures that the page mapping cannot be altered without being detected.

\subsubsection{Memory Color.} 
This field is extremely versatile and can be configured by software on each privilege level.
By choosing appropriate colors, one can segregate memory pages at runtime and also facilitate sharing across security domains.

\subsection{Solving the Challenges}\label{sec:rvas:rvasSolvingTheChallenges}
\subsubsection{Process Isolation} with \rvas' cryptographic strength could significantly enhance the security of processes inside encrypted virtual machines (\cf~\cite{AMDSevSnp}).
To solve challenge \ChalProcIso, we incorporate two tweak components:
First, privilege separation is achieved via the privilege bits in the tweak.
The CPU implicitly provides the privilege level.
Thus privileged memory is inherently separated via encryption from unprivileged software \emph{without the need for inspecting page tables}.
Second, to also isolate different processes, one can include a process identifier in the memory color field, which cryptographically separates them.

For process isolation, the memory color is chosen by the OS.
To grant it occasional access to user memory (\eg for syscall handling), one can tie the privilege level to the so-called \riscv Supervisor User Memory~(SUM) bit, which is comparable to x86 Supervisor Mode Access Prevention~(SMAP).
That is, the OS can temporarily increase the privilege level to U-mode.

\subsubsection{In-process Isolation.}
To solve the challenge \ChalSegmentMeta for segmentation, the tweak's \textit{Segment and Address Information} field can be used.
By including segmentation registers from all privilege levels, we can enforce cross-privilege policies.
We use a segment-relative address offset in the tweak such that segmentation policies can be portable between different applications, as we will show for cross-enclave shared memory.

To solve the challenge \ChalMemoryColor for memory coloring, our memory color field of the tweak facilitates a huge number of colors (\eg $2^{80}$ as opposed to 16 for Intel MPK).
Thus, \rvas makes \emph{trusted metadata storages} for memory colors (\ie tagged memory) obsolete~\cite{DBLP:conf/osdi/ZeldovichKDK08}.
Having so many unique colors available even gives brute-force resistance when used as a shared secret, as we will demonstrate.
Furthermore, \rvas can choose between memory colors from all privilege levels.

\subsubsection{Enclaves}
are the most involved isolation mechanism addressed in this paper, touching upon the challenges \ChalSegmentMeta -- \ChalMemoryEncryption.
In the following, we sketch how \rvas solves the challenges \ChalAddressTranslation and \ChalProtectedSharing.
A more detailed explanation will be given in \Cref{sec:enclaveDesign} and \Cref{sec:implementationDetails}, followed by a security evaluation in \Cref{sec:securityanalysis}.

Current enclave systems like \intel SGX~\cite{DBLP:conf/isca/McKeenABRSSS13} use trusted metadata stores, \ie the \ac{epcm}, for shadowing page table entries for each enclave page, along with an enclave identifier.
During an access, the \ac{epcm} is checked to verify that the page mapping has not been manipulated, thus ensuring the attribute invariant \IA and the spatial invariant \IS.
However, the \ac{epcm} has a few drawbacks: (1) It increases the \ac{tcb}. (2) It takes up memory. (3) The enclave's \ac{tlb} entries must be flushed during context switches~\cite{IntelSDM,DBLP:journals/iacr/CostanD16}. (4) It permits only a single owner enclave for each page, precluding flexible enclave memory sharing by design.

To overcome these limitations and solve challenge \ChalAddressTranslation, we leverage \rvas and make the \ac{epcm} obsolete:
First, we use the page table attribute field in the tweak to uphold \IA.
Second, the address and segmentation information field links between the virtual offset of the enclave and its physical page.
Moreover, we use the memory color field to label enclave pages with a unique enclave identifier.
This identifier ensures our spatial invariant \IS, since pages can only be mapped correctly to their legitimate enclave.
Note that for the relevant parts of the tweak we use M-mode privileges such that only a trusted software entity can initialize enclave pages in this particular way.
Since the temporal invariant \IT involves dynamic memory management, we will discuss it later.

To overcome the protected sharing challenge \ChalProtectedSharing, the memory color field of the tweak (\ChalMemoryColor) can be combined with an enclave-defined segment (\ChalSegmentMeta).
The segment specifies the shared memory.
The relative addressing of segments allows the enclave to choose the exact location of shared memory.
The memory color essentially comprises a shared secret established between two or more enclaves.
Only if the memory color is exactly the same, the enclaves will have the same encryption tweak and, thus, can access the shared memory.

\subsubsection{Memory Encryption}
guarantees protection against active and passive physical attacks, thus solving challenge \ChalMemoryEncryption.
For \rvas, a \ac{mee} needs to fulfill three basic properties: (1) confidentiality, authenticity, and integrity of the data, (2) offer replay protection, (3) the used cryptographic primitive must be tweakable with \tweaklength bits.
Integrity is typically ensured by storing authentication codes in a tree structure. 
The replay protection from (2) is usually done with some data structure incorporating authenticated counters.
The counters are typically fed into the encryption scheme as a tweak or nonce~\cite{DBLP:journals/iacr/CostanD16,DBLP:conf/isca/WongHC18,DBLP:conf/fpl/WernerUSSM17,DBLP:journals/jce/UnterluggauerWM19,DBLP:conf/ches/ElbazCLTSG07}.
To fulfill (3), we require a tweakable block cipher or authenticated encryption scheme with sufficiently large tweak size or associated data, such as~\cite{ascon,acorn,AEGIS128,ocb,colm}.
SGX's underlying \ac{mee} is more explicit about the used cryptographic primitives and would require changes to fulfill the third property.

\section{\ourEnclave}\label{sec:enclaveDesign}

This section introduces our \ourEnclave enclave architecture to highlight the most complex use case of \rvas. 
We first give an overview of the design, then describe the threat model, and finally, explain how we enable secure enclaves on top of the authenticryption shield.

\subsection{Overview}
\begin{figure}
	\includegraphics[width=0.8\columnwidth]{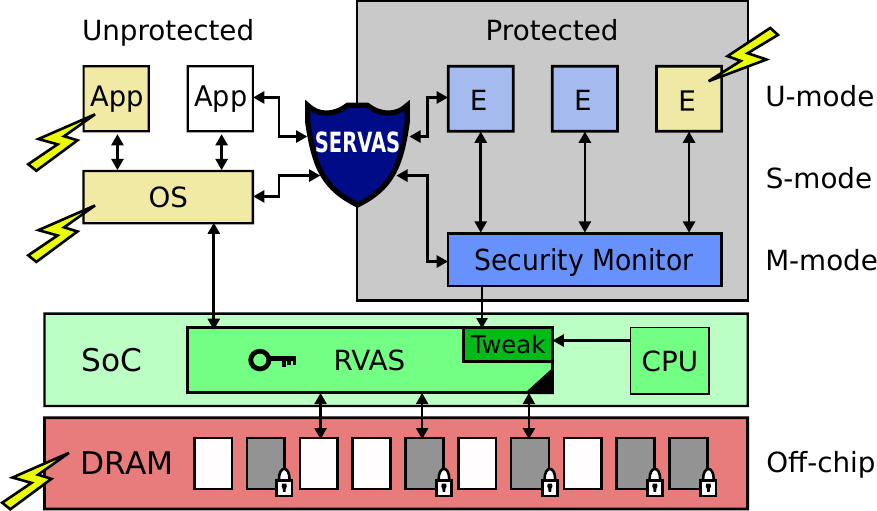}
	\centering
	\captionof{figure}{Overview of \ourEnclave: Thunderbolts show the physical and software attack surface. \textit{App.} denotes unprotected user-mode applications and \textit{E} denotes enclaves. \rvas encrypts and authenticates pages in the untrusted DRAM. Pages with a lock symbol belong to enclaves.}
	\label{fig:coarseoverview}
\end{figure}

\ourEnclave is an innovative and highly flexible isolation mechanism for securely executing enclaves.
As shown in \Cref{fig:coarseoverview},
\ourEnclave consists of the \acf{rvas} and a software \acf{sm} that manages the whole enclave's lifecycle.

Our \ourEnclave system follows SGX's design choices to keep a minimal \ac{tcb} while simultaneously avoiding the drawbacks of large trusted metadata storages (\ie the \ac{epcm}). 
Instead, we feed the relevant security metadata into the \rvas tweak.
By carefully controlling the encryption tweak, \ourEnclave maintains cryptographic segregation of various security domains.
Additionally, \ourEnclave also enables dynamic enclave memory and natively supports secure sharing of enclave memory, avoiding costly software-based encryption~\cite{anati2013innovative}.

In our design, the \acf{sm} is the trusted intermediary that acts as a universal entry and exit point for enclaves.
Hence, any interactions between untrusted software components and enclaves are safeguarded by the \ourSM.
Moreover, the \ac{sm} is responsible for loading enclaves, initializing their memory pages, handling syscalls and interrupts, shared memory, and swapping.
To perform its tasks, the \ourSM is capable of controlling most parts of the tweak.

In the following, we discuss our threat model and show how \ac{rvas} helps us build an enclave architecture.

\subsection{Threat Model}
\ourEnclave protects enclaves that consist of security-sensitive code and perform operations on sensitive data.
Our threat model is mostly in line with Intel SGX~\cite{DBLP:conf/isca/McKeenABRSSS13,DBLP:journals/iacr/CostanD16} and considers a powerful, privileged software and physical attacker.
Software running outside the enclave (\ie the \ac{os} and user applications) is viewed as untrusted and might be subject to compromise.
We assume the adversary has full control of the \ac{os} and can launch rogue applications and enclaves.
Unlike SGX, the adversary can tamper with the memory mapping of enclaves (\eg \pts) and also exercise unrestricted software access to the physical DRAM. 

Our \acf{tcb} comprises both software and hardware components.
On the software side, we only trust the enclaves themselves and a small \acf{sm} for enclave management.
The enclave developer is responsible for adequately implementing enclaves. 
Vulnerabilities in the enclave itself cannot be prevented by \ourEnclave~\cite{DBLP:conf/uss/LeeJJKCCKPK17,DBLP:conf/uss/BiondoCDFS18}.
Also, malicious enclaves~\cite{DBLP:conf/dimva/0001WG19} are outside our threat model but could be addressed by orthogonal defenses~\cite{DBLP:conf/raid/WeiserMSG19}.

The \ac{sm} is running in \riscv M-mode and can also be viewed as an integral part of our CPU hardware.
For example, Intel SGX implements its enclave management via CPU microcode~\cite{DBLP:journals/iacr/CostanD16}. 
We assume that the \ourSM is stored on a trusted on-chip ROM or verified as part of a secure boot process~\cite{riscv2020secureboot}.

On the hardware side, the \ac{tcb} covers our \acf{soc}, including the CPU core and \ac{rvas}. 
Anything outside the \ac{soc} is untrusted.
In particular, the attacker can tamper with the DRAM and mount bus probing or cold-boot attacks~\cite{DBLP:conf/IEEEares/LindenlaufHS15}. 
\ourEnclave effectively removes CPU components involved in the page mapping and address translation from the \ac{tcb}:
unlike SGX, which needs to store trusted metadata in the EPCM~\cite{DBLP:journals/iacr/CostanD16}, \ourEnclave avoids having an EPCM, thus slightly decreasing the \ac{tcb} complexity of our \ac{soc}.

Denial-of-service attacks are outside of our threat model.
It is up to the \ac{os} and the applications to invoke an enclave.

Side-channel attacks, whether performed in software or in hardware, are an orthogonal challenge, for which plenty of literature is available that could also be applied to \ourEnclave.
We will discuss side-channel security in \Cref{sec:securityanalysis}.
Fault attacks on the encrypted DRAM~\cite{DBLP:conf/sosp/JangLLK17} are detected by \rvas' authenticated encryption, while fault attacks on the \ac{soc}~\cite{DBLP:conf/sp/MurdockOGBGP20,DBLP:journals/corr/abs-1912-04870} are outside our scope.

\subsection{Building Enclaves}\label{section:building_enclaves}

\ourEnclave enclaves are built on top of \rvas with the assistance of our \acf{sm}.
The \ourSM uses the \rvas \ac{isa} extension to configure the tweak and enforce our \textit{spatial- } (\IS),  \textit{temporal-} (\IT), and \textit{attribute invariant} (\IA), as specified in \Cref{sec:enclave_challenges} over the lifecycle of an enclave.

We ensure that enclaves are separated even across otherwise identically structured processes, as follows:
We include address information, to confine enclave execution, such as page mappings, page permissions and virtual address range checks as part of \rvas' segmentation and address information, privilege level and page table attributes fields. 
We further embed page ownership information as part of the memory color.

In total, we define four different page types for specific use-cases in \ourEnclave:
\ptregular denotes general-purpose memory for private enclave code and data. 
It must reside in an enclave-specific virtual memory range.
\ptshenclave enables different instances of the same enclave to deduplicate read- and execute-only pages.
Thus, it can reduce the memory overhead significantly.
We also define \ptshm, a page type that allows enclaves to set up secure cross-enclave data sharing by providing a shared secret before the access.
Finally, the \ac{sm} exclusively uses \ptmonitor pages to store information about the loaded enclaves in memory. 
This page is referenced when switching to an enclave to load the enclave's state or save its register content if interrupted (\eg by a timer interrupt).
In the following, we present how the page types link to different types of memory.

\subsubsection{Secure Static Page Mapping.}
To keep \IS and assure a secure virtual-to-physical page mapping of any enclave-related page, we initialize the page with mapping information in the tweak. 
Moreover, segment and address information, and page table attributes assure that the mapping is only valid for a specific virtual address and well-specified access permissions.

For private enclave memory, we use a unique \acf{rtid} of the enclave as the tweak's memory color to bind the page to its enclave.
This single ownership principle helps to distinguish enclave instances and satisfies the \textit{temporal invariant} \IT.
\ptmonitor memory solely belongs to the \ourSM, which we enforce by including the privilege mode in the tweak.

To share enclave code or read-only enclave data and to help reduce memory load and \ac{tlb} pressure, developers can mark pages as \ptshenclave.
In this case, a so-called \acf{encid} substitutes the \rtid.
The \encid uniquely identifies an enclave codebase or binary via a cryptographic authentication code.
Since multiple owners exist, \ptshenclave pages need to be read-only to enforce \IT.

\subsubsection{Secure Dynamic Page Mapping.}
\ourEnclave enclaves may use dynamic memory, which has been allocated by the host user-mode application.
To do so, the enclaves ask the \sm to cryptographically initialize the memory as before.

While static enclave mappings are secured by the \ourSM, dynamic mappings that change at enclave runtime demand special care.
To prevent double mapping attacks and enforce the security invariant \IT, we first require the enclave (runtime) to keep track of all of its valid page mappings in a private bitmap (similar to SGX~\cite{mckeen2016intel}).
Thus, when the enclave receives new memory from the host, it can consult its private bitmap and refuse this memory if it is already in use.
Second, if an enclave releases dynamic memory, it explicitly invokes the \sm, which invalidates the page by destroying its integrity.
This invocation prevents use-after-free scenarios and upholds \IT.
Note that the private bitmap only needs to enumerate pages \textit{inside} the enclave's address range.

\subsubsection{Enclave Data Sharing.}
\ourEnclave introduces a novel concept of enclave data sharing.
\ptshm memory is writable and can be used for data exchange at native speed (\ie without copying or re-encryption~\cite{anati2013innovative}).
As with dynamic memory, the \ourSM performs the setup of shared data pages for enclaves and requires that the respective enclaves acknowledge any mapped page. 
\ptshm memory is identified by a shared secret that is directly managed by the enclaves.
The \ourSM can further assist in establishing a shared secret between the two enclaves by acting as a trusted entity attesting the respective enclaves to each other.

Upholding our security invariants for data sharing is critical and highlights the versatility of our \rvas design.
Enclaves can enforce \IS by simply keeping the shared secret confidential, that is, by securely generating and distributing shared secrets with the help of \sm.
Data sharing also seamlessly scales to multiple enclaves.
A user-mode range register points to the desired shared virtual memory range to uphold \IT, by preventing double mapping attacks (\eg aliasing \ptshm with existing \ptregular pages)

\subsubsection{Enclave Life Cycle.}
The \acf{sm} is responsible for the whole enclave life cycle and is loaded as part of a secure boot procedure. 
This procedure is part of the ongoing work of the \riscv groups~\cite{riscv2020secureboot}. 
In line with other enclave designs~\cite{DBLP:conf/eurosys/LeeKSAS20}, our \ourSM provides an API for managing all aspects of an enclave's lifetime: loading, entering, exiting, interruption, managing, and initializing memory for the enclave's code, stack, dynamic memory, shared memory and swapping.
Furthermore, it provides functionality for local attestation and sealing.
To perform operations involving the enclave's virtual memory, the \ourSM can override most parts of the tweak that is passed to \rvas.
Thus, the \ourSM can initialize enclave memory as if the enclave itself did it.
For the detailed API description, we defer to \Cref{app:sm_api}.

Enclaves can be distributed as encrypted binaries to prevent the host application from accessing the enclave's code or data.
A user-mode application can choose to load and run an enclave within its own virtual address space.
Enclave binaries are then decrypted (if necessary) and authenticated within the \sm, where their code and data are initialized with the page type \ptregular or \ptshenclave.
A loaded enclave can be entered at its single developer-specified entry point.

For encrypted enclaves and use-cases that require deployment on specific systems, an AEAD encryption scheme (\eg \ascon~\cite{ascon}) with a developer key derived from a per-CPU key can be used. 
Encryption eliminates the need for code obfuscation \cite{DBLP:conf/ccs/LinnD03} to prevent theft of intellectual property.
Additionally, secrets can be directly embedded in the enclave code.
The decryption happens in software within the \sm. 
Loading an enclave also authenticates it and generates a load-time hash forming the \ac{encid}, achieving load-time attestation.
To store enclave secrets, the \sm provides a sealing functionality. 

\subsubsection{Swapping}
is essential to handle out-of-memory situations.
Enclave systems like \intel SGX can only use a limited physical memory.
In contrast, \ourEnclave allows using all of the available memory for enclaves, making out-of-memory situations happen less frequently.
Nevertheless, for real-world workloads, a swapping mechanism is key. 
Swapping of enclave pages requires interaction between the \ac{os} and the \sm.
To ensure the \sm's correct operation and maintain our security invariants, we exclude \ptmonitor and \ptshm pages from being swapped. 
When the \ac{os} selects an enclave page to be swapped out, it provides a temporary page to the \ourSM to safely move the page to.
The \sm will then re-encrypt the page-to-swap to this temporary page by using an AEAD encryption scheme.

Afterward, to uphold \IT, the \sm invalidates the original physical enclave page. 
The involved metadata (\ie AEAD-tag, used nonce, virtual address, range information, and page permissions) are  saved on a per-enclave \ptmonitor page. 
This metadata ensures that only the one specific version of the page can later be swapped in and prevents any roll-back attacks. 
The \ac{os} can then write the temporary page safely to the disk and repurpose it for other applications.

\section{\ourEnclave Implementation Details}\label{sec:implementationDetails}
\begin{figure*}
	\centering
	\begin{subfigure}[t]{0.4\textwidth}
		\centering
		\includegraphics[width=\textwidth]{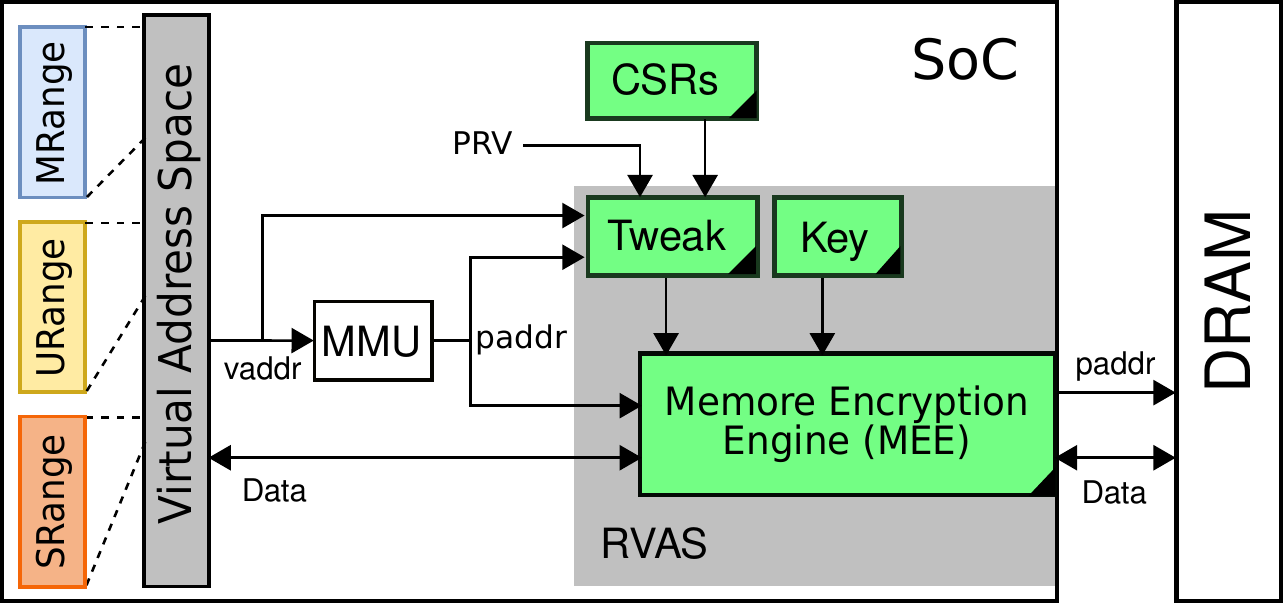}
		\caption{Blocks with black corners show components that are part of \rvas.}
		\label{fig:overview}
	\end{subfigure}
	\hfill
	\begin{subfigure}[t]{0.55\textwidth}
		\centering
		\includegraphics[width=\textwidth]{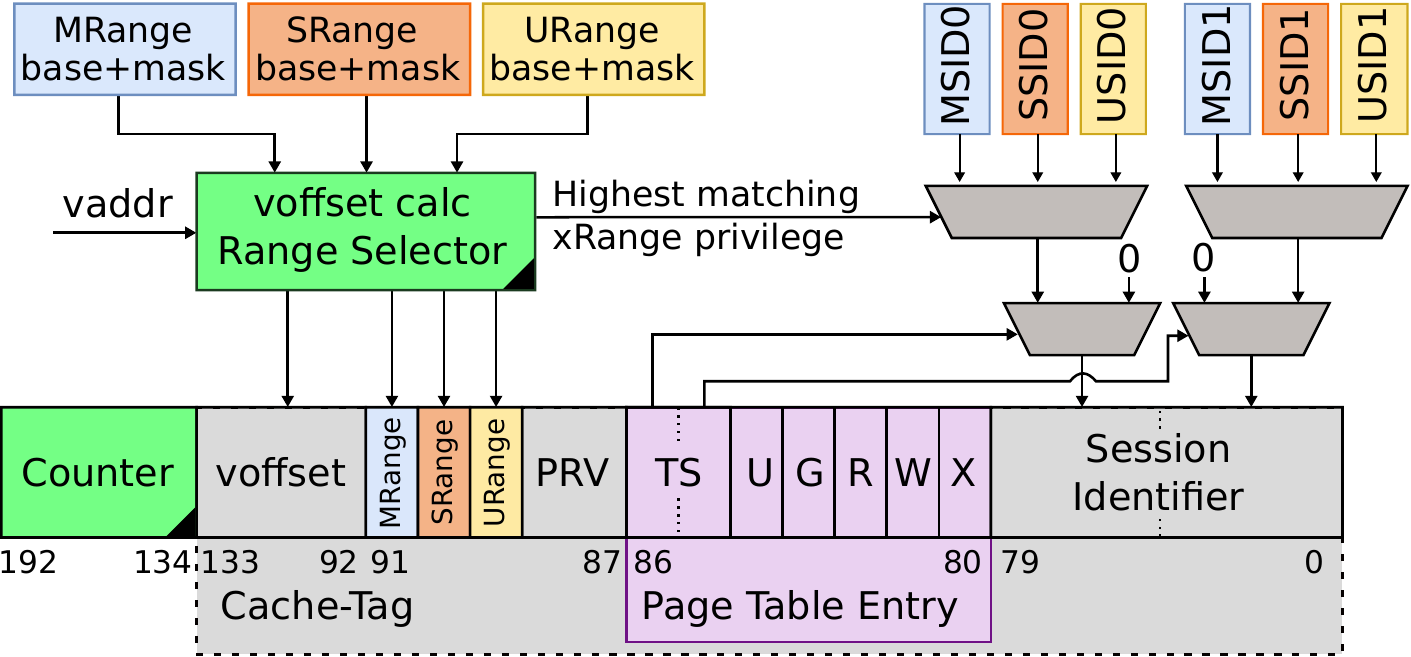}
		\caption{Composition of the \rvas tweak fed into the \ac{mee}. The counter is part of \rvas's replay protection. \ptrsw denotes our tweak-select bits stored in the \ac{pte}.}
		\label{fig:tweakStructure}
	\end{subfigure}
	\caption{Overview of \rvas and the tweak construction for the \ac{mee}. PRV denotes the current privilege mode.}
	\label{fig:rvas}
\end{figure*}

In this section, we detail the implementation and parameterizations of our prototype implementation of \ourEnclave .
An overview of how \ourEnclave uses \rvas for this matter can be seen in \Cref{fig:overview}.
First, we highlight the required ISA extensions and then show how the tweak for \rvas  is assembled.
Second, we detail the new page types introduced by \ourEnclave and describe the security monitor managing the enclaves.
Finally, we explain how a tweak cache can reduce the hardware footprint of \ourEnclave.

\subsection{Instruction Set Extension Details}
Realizing \ourEnclave only requires minimal changes.
First, we extend the existing \riscv ISA with additional \acp{csr} to set the tweak of \rvas in software.
Second, we add a so-called \authenticationExceptionAsText, which is raised by the \rvas whenever the decryption fails with an integrity check error during a read, write, or fetch operation.
Handling this exception in our \sm is a key ingredient for the security of \ourEnclave.

The additional \acf{csr} for the machine- (M), supervisor- (S), and user- (U) mode that we refer to as \Mrange, \Srange, and \Urange, can be seen in \Cref{fig:tweakStructure}.
The ranges consist of a base address and a size and specify regions in the virtual address space, which allows us to differentiate between memory accesses.
Any access to an \Xrange must pass \rvas. 
The \Mrange is used by the \ourSM to declare the memory of the enclave.
The \Srange and \Urange give enclaves control over additional encrypted ranges for enclave-specific needs like shared memory.

To include software-controllable \acfp{sid} in the tweak, we add two \acp{csr} for each privilege mode, namely \texttt{x{\sid}0} and \texttt{x{\sid}1}.
To decide which of the two \sid registers is used for the tweak, we repurpose two additional selector bits in the currently unused most significant bits part of the \acp{pte} as tweak-select (\ptrsw).

\subsubsection{Tweak Override.}
Moreover, we define special load- and store-tweak registers, which are only accessible in machine mode to our \ourSM. 
These registers can be used to override any tweak parameters used by \rvas and also disable them, except for the \rvas-managed integrity counters.
This ``tweak override'' allows the \ourSM to cryptographically initialize a page in lieu of the enclave without trusting the \ac{os}-supplied page mapping. 
Any further accesses by the lesser-privileged modes must adhere to the same tweak used for initialization, which is cryptographically enforced by \rvas.

Lastly, we assume an additional per-CPU key for encrypted loading of enclave binaries, which is fused into the chip and only accessible to our \sm via \acp{csr}.

\subsection{Tweak}
\label{sec:tweakdesign}
\rvas uses the tweak to incorporate additional information about the CPU state into its \ac{mee}, allowing us to achieve a strong software-defined cryptographic separation between enclaves. 
As shown in \Cref{fig:tweakStructure}, the tweak incorporates integrity counters, page mapping information, privilege information, range checks, and software-controllable session identifiers.
For \ourEnclave, we propose to use \rvas with a tweak size of 192 bits in total.
It is structured as follows:

\subsubsection{Counter.} Similar to SGX's 56-bit counters, we reserve 58 bits of the tweak for the \textit{integrity counter} to guarantee freshness and protect against replay attacks \cite{ARMMemEncPresentation,DBLP:journals/jce/UnterluggauerWM19,DBLP:journals/iacr/CostanD16}.
The counters are managed inside \rvas and not exposed to software.

\subsubsection{xRange.} We use 3 bits to encode whether the accessed address is within \Urange, \Srange, or \Mrange, respectively. 
These bits constitute the bitmap in the \textit{segment and address information} field of the \rvas tweak design.
The range boundaries can be set from their respective privilege levels via the \Xrange registers.
If an address matches an \Xrange register, its tweak bit is set to one.
This matching enforces a strong domain separation between the specified memory ranges. 
The rightmost \Xrange bit set to one determines the \texttt{voffset} calculation and the choice of the \texttt{x{\sid}} registers (\eg \Urange has precedence).

\subsubsection{voffset} is the virtual address offset and complements \Xrange as part of \rvas' \textit{segment and address information} field.  
The voffset is computed from the base address of the rightmost matching \Xrange register at cache line granularity.
For 48-bit virtual addresses~\cite{riscvPrivISA} and \SI{64}{\byte} cache lines~\cite{DBLP:journals/iacr/CostanD16}, the bit field is $48 - log_2(64) = 42$ bits.

\subsubsection{PRV} encodes the current privilege mode of the CPU in two bits.
Having these bits in the tweak gives us a strong privilege separation.
This field corresponds to the \textit{privilege level} field in the \rvas tweak design.

\subsubsection{\ac{pte}.} 
Since the untrusted \ac{os} manages the page mapping, we include seven bits from the \ac{pte} in the tweak. 
These \ac{pte} bits includes the U bit, deciding whether a page is accessible in user mode, 
the G bit, defining if it is a global mapping, and 
the three RWX bits, stating the read, write and execute privileges of the page.
We also include two of the software-defined reserved tweak select \ptrsw bits in the tweak for selecting which of the \sid registers are used.
The \ac{pte} maps to the \textit{page table attributes} field of the \rvas tweak design.

\subsubsection{\ac{sid}.} We allocate \SI{80}{\bit} for the \acf{sid}, which corresponds to the \textit{memory color} field in the \rvas tweak design.
The rightmost matching \texttt{x{\sid}} register determines whether \texttt{M{\sid}}, \texttt{S{\sid}}, or \texttt{U{\sid}} is used.
\ptrsw determines whether one or both  \texttt{x{\sid}0} and \texttt{x{\sid}1} registers are used.
If both are used (\ie \ptrsw = \texttt{11b}), the resulting \ac{sid} is truncated to \SI{80}{\bit}.

\subsection{Page types}
\ourEnclave defines various page types, which use a specific combination of the tweak components, as seen in \Cref{tab:tweak_composition}.

\subsubsection{\ptunprotected}
marks any untrusted page. 
This page type is located outside of any of the \Xrange regions and is accessible from any privilege level with the page permission configured in the \ac{pte}.
While \rvas can also encrypt \ptunprotected memory, one could bypass encryption to achieve native performance for non-enclave applications. 

\subsubsection{\ptregular}
denotes pages in the \Mrange and is intended to be used with a single enclave instance from user mode.
The pages can have any combination of page permissions, as specified by the \ac{pte}.
The \ptrsw bits specify the use of the \msidreg{0} register that holds the unique, \ourSM-defined, \rtid value for the enclave instance.

\subsubsection{\ptshenclave}
can be used to share non-writable pages between different instances of the same enclave.
This sharing allows to reduce memory requirements and \ac{tlb} pressure significantly.
This page type adheres to the \Mrange and uses the \msidreg{1} \ac{csr}, which holds a unique \acf{encid} for the loaded binary.

\begin{table}[b]
	\setlength{\tabcolsep}{4pt}
	\setlength{\aboverulesep}{0pt}
	\setlength{\belowrulesep}{0pt}
	\caption{Tweak decision table: $\bullet$ denotes an arbitrary value, ugrwx and \ptrsw correspond to the respective \ac{pte} bits, and PRV holds the current privilege level.}
	\label{tab:tweak_composition}
	\centering
	\resizebox{\columnwidth}{!}{%
		\begin{tabu}{cccccc|cl}
			\toprule
			\Mrange & \Srange & \Urange & PRV & PTE & \ptrsw & \acs{sid} & Label \\			\midrule
			0 & 0 & 0 & $\bullet$ & $\bullet$ & $\bullet$ & $\bullet$ & \ptunprotected \\
			1 & 0 & 0 & U & $\bullet$ & 01 & \msidreg{0} & \ptregular \\
			1 & 0 & 0 & U & !W & 10 & \msidreg{1} & \ptshenclave \\
			0 & 0 & 1 & U & !X & 11 & \usidreg{0+1} & \ptshm \\
			$\bullet$ & $\bullet$ & $\bullet$ & M & rw & $\bullet$ & 0 & \ptmonitor \\
			\bottomrule
		\end{tabu}
	}
\end{table}

\subsubsection{\ptshm}
can be used to safely share data between instances of enclaves, also between different enclaves.
Such pages can only reside in the \Urange, which the \ourSM assures at initialization time.
For memory inside the \Urange, \rvas feeds the virtual address \textit{offset} relative to the \Urange base into the tweak.
This relative offset ensures cross-address-space accessibility of the shared memory.
To counteract runtime attacks, \ptshm pages can never be executable.

The \ptrsw bits from the \pt indicate that both \usidreg{} registers are used for the \sid.
Enclaves load the shared secret to the \usidreg{0} and \usidreg{1} registers before accessing the shared memory region.
With the help of the \ourSM, two or more enclaves can agree on an 80-bit shared secret, which separates different shared-memory regions from each other.
By specifying \Urange appropriately, an enclave can constrain shared memory to a particular region and, thus, prevent accidental double mapping attacks.

\subsubsection{\ptmonitor}
denotes pages used by the \ourSM to store metadata for each enclave and thread.
\ptmonitor must only be accessible by the \ourSM, \ie in the M mode.

\subsection{\acf{sm}}\label{section:impl_sm}
The \sm manages enclaves and their transitions between the different privilege modes. 
It runs with machine-mode privileges, stores its tiny code base and the stack on-chip, and is loaded as part of a secure boot process.
Both areas are protected using a \ac{pmp} such that only the \ac{sm} can access it.

\subsubsection{Stateless \sm.}
In principle, the \sm can run completely stateless and only requires a small (approx. 1KiB) stack during operation.
No static state needs to be kept between \sm invocations since all enclave management data is stored inside dynamically allocated \ptmonitor pages managed by the operating system.
The \sm also allocates a unique \rtid to each enclave instance.
Our prototype currently stores the previously allocated 64-bit \rtid in memory and increments it for each new enclave to guarantee uniqueness.
To remove this tiny 64-bit state, one could simply sample the \rtid from one of the \riscv hardware performance counters, \eg the elapsed CPU cycles \texttt{mcycle} or other CPU events \texttt{mhpmcounter}.
The \sm only needs to ensure that they are monotonically counting CPU-global events that assuredly occur between two enclave creations (e.g., memory accesses).
For a worst-case estimate, we use \texttt{mcycle} incrementing at 5GHz speed.
Thus, it will overflow the 64-bit range only after 117\,years of continuous operation.
After the device reboots, no enclave is running, and we can sample the \rtid freshly.

\subsubsection{Enclave metadata} is stored in OS-managed but RVAS-protected per-enclave and per-thread \ptmonitor pages.
In order to access \ptmonitor pages, the \sm leverages the tweak override feature to ensure its exclusive access.

\subsubsection{Other pages} are similarly initialized by the \sm using the tweak override to force their page type (\eg PTE bits, privilege mode, and other tweak-associated \acp{csr} such as \Xrange and \texttt{ xSID\{0,1\}}).

\subsubsection{API} The \ourSM provides an API that works like syscalls but trap into the \sm instead.
The API allows to manage the lifecycle of the enclave or interact with the \sm, \ie creating, entering, exiting, page preparation, page destroying, in-place re-encryption, and acquiring a sealing key.
A detailed description of the API can found in \Cref{app:sm_api}.

\subsubsection{\ac{sm} Prototype.}
Our prototype implements all of the above API functionality. 
In total, our \sm has a very small code size of \locsmcomplete~\acp{loc}.
Of this, \locascon~\acp{loc} are taken up by the \ascon implementation, which is used for loading encrypted enclaves, and deriving the sealing key. 

\subsection{Caching}\label{sec:tweakCache}
For \ourEnclave, we extend the cache to store the tweaks next to each cache line, referred to as the \textit{inline variant}.
This extension allows us to compare the tweak stored in the cache line to the tweak for the access and ensures that the entire tweak can be reconstructed for any write-back operations.
Our tweak design described in ~\Cref{sec:tweakdesign} consists of \SI{192}{\bit}.
However, as the \ac{mee} of \rvas manages the \SI{58}{\bit} integrity counter, we only need to store the remaining $b_{\ourEnclave}$ \SI{134}{\bit} tweak in our extended cache.
These remaining bits $b$ consist of $b_{xrange}$ bit for \Urange, \Srange, and \Mrange, $b_{pte}$ bits from the \ac{pte}, $b_{prv}$ bits from the privilege level, $b_{sid}$ bits for the \ac{sid} and $b_{voffset}$ bits for the voffset.
Each of the $N_{cache}$ cache lines in the data- and instruction cache are augmented with the tweak.
Note, that $N_{cache}$ might vary between the two caches.
Therefore, the required storage $S_{Total}$ can be calculated as the sum of additionally stored bits in the data- ($S_{Data}$) and instruction cache ($S_{Instr}$).
%\resizebox{.9\linewidth}{!}{
\begin{minipage}{\linewidth}
	\begin{align}
	b_{\ourEnclave} &= b_{voffset} + b_{xrange} + b_{prv} + b_{pte} + b_{sid} \label{equ:servastweak}\\ 
	S_{Data} &= S_{Instr} =b_{SERVAS} \cdot N_{cache} \\
	S_{Total} &= S_{Data} + S_{Instr}
	\end{align}
\end{minipage}
%}

\subsubsection{Cache Optimization.}
\begin{figure}
	\includegraphics[width=0.9\columnwidth]{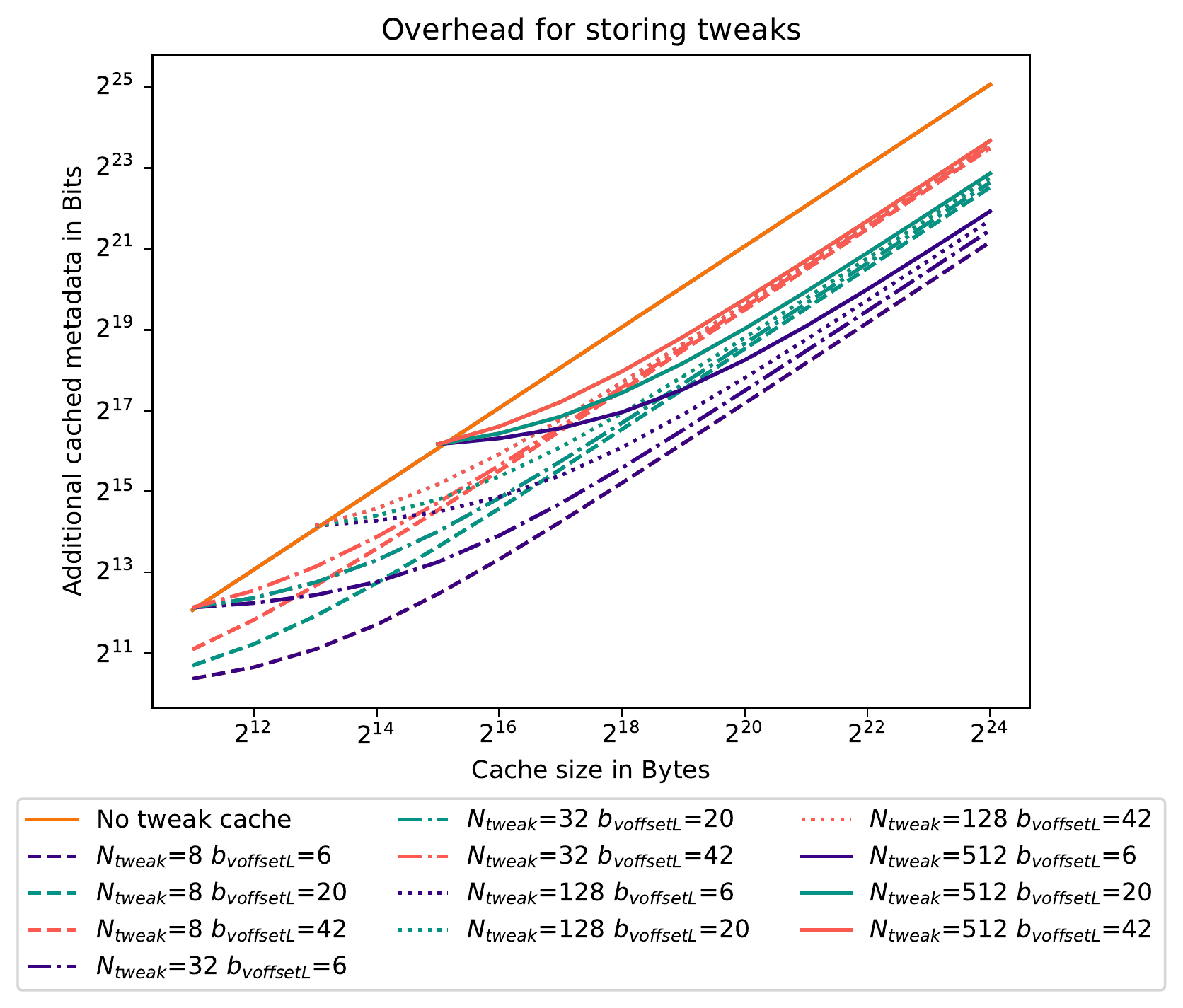}
	\vspace{-0.5em}	
	\centering
	\captionof{figure}{Overhead of the tweak cache in bits compared to the inline variant for a different number of tweak cache entries $N_{tweak}$. $b_{voffsetL}$ denotes whether 6, 20 or 42 bits of the voffset are stored in the main cache. The cache line size is \SI{512}{\bit}.}
	\label{fig:tweakcache_overhead}
\end{figure}
For many real-world scenarios, we only have a limited number of different tweaks, which could be deduplicated in a separate \ac{tc}~\cite{DBLP:conf/iccd/JoannouWKMBXWCR17}.
This deduplication would help to significantly shrink the tag size in the caches and, thus, the area overhead.
For simplicity, we refer to the data and instruction caches as the \textit{main caches}.
For our design, such a \ac{tc} could be implemented using an index-based indirection in the main cache and a separate \ac{tc}, whereas the index is stored along the cache line to link the cache and \ac{tc}.
Appropriately sizing a cache is generally tightly coupled to the system's expected workload. 
In the case of the \ac{tc}, it also depends on the expected number of enclaves running in parallel.
Therefore, a precise parameterization of such a cache exceeds the scope of this paper.
However, in the following, we want to discuss the parameters that can be used to tune the cache and their effects.

First, insertion into the \ac{tc} can be handled by using a non-linear function like a lightweight cryptographic hash- or permutation-based function to derive the set index from the tweak pseudorandomly, similar to \cite{DBLP:conf/uss/WernerUG0GM19,DBLP:conf/parle/SeznecB93}.
This index generation makes finding the correct index for already present tweaks in the \ac{tc} efficient.
Moreover, since the \ac{tc} index can be computed in parallel to the main cache lookup, the latency can be mostly hidden by choosing a primitive matching the cache latency.
Therefore, only one additional cycle for the tweak comparison may be required.
For tweak replacements in the \ac{tc}, all associated cache lines, \ie all cache lines with that specific index, need to be flushed.
Each time a new tweak is inserted into the cache, there is a probability that another cache line is evicted. 
To reduce the probability of two or more tweaks taking turns in evicting each other, the cache can be: (1) made larger or (2) be split into multiple ways~\cite{DBLP:journals/tc/HillS89}. A more detailed analysis and simulation can be found in \Cref{app:eviction_probability}.

Next, we define a range in virtual memory that uses the same tweak (except for $b_{voffsetL}$ bits of the voffset) as a tweak zone that each enclave has a number of.
In general, the main cache  stores more cache lines than the \ac{tc} stores tweaks.
Hence, if the expected size of tweak zones is relatively small, additional $b_{voffsetH}$ bits of the upper voffset can be stored more cheaply in the \ac{tc}, than the main caches.

Moreover, to size the \ac{tc} to improve upon the inline variant, we must take two main constraints into account: (1) additional ways in the \ac{tc} require additional parallel comparator logic and (2) the stored bits in the \ac{tc} and the additional bits in the main cache must be smaller than the inline variant.
For the inline variant, the tweak is compared in parallel to the cache tag.
Hence, to handle (1) there should be fewer ways in the \ac{tc} than in the main cache to reduce the overhead of comparators.
Addressing (2) is more difficult since it relies on assumptions on the size of tweak zones, the size of the main caches and the \ac{tc}'s desired size.
We need to reconstruct the tweak $b_{SERVAS}$ from \Cref{equ:servastweak}, but the voffset can be split into an upper $b_{voffsetH}$ and a lower part $b_{voffsetL}$, with the latter being stored along the cache line in the main cache.
Additionally, $b_{tweakidx}$ bits are required in the main cache to link the correct entry in the \ac{tc}, the width of which depends on the number of entries $N_{tweak}$ in the \ac{tc}.
The remaining bits stem from the $N_{tweak}$ entries in the \ac{tc} times the rest of the tweak and a valid bit $b_{valid}$.
The overhead in each cache can then be aggregated into $S_{TotalOpt}$:

\begin{align}
	b_{voffset} &= b_{voffsetH} + b_{voffsetL} = \text{const.} \\
	b_{tweakidx} &= \log_{2}(N_{tweak}) \\
	S_{DataOpt} &= S_{InstrOpt} = (b_{voffsetL} + b_{tweakidx}) \cdot N_{cache} \\
	S_{tweakcache} &= (b_{valid} + b_{\ourEnclave} - b_{voffsetL}) \cdot N_{tweak} \\
	S_{TotalOpt} &= S_{DataOpt} + S_{InstrOpt} + S_{tweak}
\end{align}

We evaluate the number of additional bits required as a function of the main cache size and \SI{512}{\bit} cache lines in \Cref{fig:tweakcache_overhead}. 
The graph shows the number of additional cached tweak bits for the inline variant as the top line. 
The other lines represent \ac{tc} configurations that differ in the number of tweak entries $N_{tweak}$ and the split between $b_{voffsetL}$ and $b_{voffsetH}$.
One can observe that the break-even point for each $N_{tweak}$ is when the \ac{tc} has as many tweaks as the main cache has cache lines.
After that, the fewer bits are stored in the increasingly large main cache, \ie the smaller the $b_{voffsetL}$, the smaller this overhead becomes.
Additionally, with larger caches the $b_{voffsetL}$ quickly becomes the dominating factor and clusters them into groups.
For the same reason $b_{tweakidx}$ also contributes to this, but due to its smaller size, the effect is less pronounced.

\subsection{Encryption Bypass Optimization}
Our prototype implementation encrypts the whole system's physical memory.
However, in the future we intent to make it configurable, such that the \ac{mee} only encrypts pages that require this protection, \eg enclaves.
Hence, \ourEnclave can also be used as an extension for enclaves only.
This variant can be achieved with some logic in \rvas that uses information from the \Xrange registers to decide if a request has to go through the \ac{mee} or access the memory directly.
Some \acp{mee} like \memsec read and verify the integrity of data before new data is written to a location.
In a bypass implementation, this verification would cause issues when initialization enclave pages since the \ac{mee} attempts to verify the integrity of unprotected memory locations.
However, when initializing encrypted pages, the SM is already able to override any supplied tweak information.
One would need to simply extend this initialization mechanism by intermittently ignoring any integrity violations.

A limitation of the performance bypass is that the inherent overhead of the integrity protection trees introduced by the \ac{mee} persists.
While sparse integrity trees could address this problem, no open-source memory encryption schemes with sparse integrity trees have been proposed to the best of our knowledge.

\section{Security Analysis}\label{sec:securityanalysis}
In this section, we analyze the security of \ourEnclave and its interaction with \rvas.
\rvas cryptographically enforces the tweak (\cf \Cref{sec:security_properties}) and solves the challenges of memory isolation \ChalProcIso-\ChalMemoryEncryption.
Moreover, only the trusted \ourSM can override the tweak, effectively preventing forgery, thus, enforcing the domain separation.
However, attacks might aim at breaking the \textit{spatial-} (\IS) and \textit{temporal invariant} (\IT).
In the following, we discuss how \ourEnclave upholds these variants in a number of attack scenarios.

\subsection{Side-channel Attacks and Defenses}
Our threat model explicitly excludes side-channel attacks to adopt \intel SGX's threat model.
Additionally, we see micro-architectural attacks as specific to the underlying hardware implementation of the CPU which needs to be fixed for enclave and non-enclave code alike, hence, is out of scope of this paper.
However, in the following, we want to briefly discuss several attacks a system equipped with \ourEnclave might encounter.

\subsubsection{Physical Attacks} such as differential power analysis~\cite{DBLP:conf/crypto/KocherJJ99} could break
\rvas.
\rvas can use more expensive memory encryption schemes such as MEAS~\cite{DBLP:journals/jce/UnterluggauerWM19} to protect from physical attacks on the memory encryption. 

\subsubsection{Enclave Shared Memory} opens a harmless side-channel, allowing Flush+Reload attacks on the shared memory.
For data sharing, enclaves need to trust each other anyways and can access the shared memory directly rather than via a side-channel.
For code pages, we ensure that only instances of the same enclave can optionally deduplicate code.

\subsubsection{Controlled-channel Attacks}~\cite{DBLP:conf/sp/XuCP15}
extract side-channel information from an application's inability to access certain memory locations, \eg by leveraging the \ac{os}'s ability to keep control over page tables, thus, unmapping pages to trace accesses.
Attacks have been shown on \intel SGX Enclaves (SGXStep~\cite{DBLP:conf/sosp/BulckPS17}, CopyCat~\cite{DBLP:conf/uss/MoghimiBHPS20}).
For \ourEnclave, these types of attacks also apply.
However, provably secure interruptibility~\cite{DBLP:conf/csfw/BusiNBGDMP20} and constant-time code can be used as a mitigation technique

\subsubsection{Cache Attacks}
Modern CPUs involve many microarchitectural buffers, all of which could present a side-channel.
To close them in software, one can flush core-local buffers when entering or leaving the enclave~\cite{DBLP:journals/corr/abs-2005-02193} or use constant-time code.
For securing cross-core buffers such as last-level caches, many partitioning schemes~\cite{CATPerformance2015,DBLP:conf/uss/DessoukyFS20} and randomization-based approaches~\cite{%
	DBLP:conf/uss/WernerUG0GM19,%          %Scattercache
	DBLP:conf/micro/Qureshi18,%            %CAESAR
	DBLP:journals/corr/abs-2009-09090,%    %Mirage
	DBLP:conf/ndss/TanZB020} %               %Phantomcache
exist to provide protection on a hardware level.

\subsubsection{Transient Execution Attacks}~\cite{DBLP:conf/sp/KocherHFGGHHLM019,DBLP:conf/uss/CanellaB0LBOPEG19}
present a threat for many modern high-performance CPUs in both enclave and non-enclave execution mode.
To combat this threat, we can use techniques and additional instruction as proposed by MI6~\cite{DBLP:conf/micro/BourgeatLWZAD19}. 
Furthermore, Wistoff et al. \cite{DBLP:journals/corr/abs-2005-02193} have shown how transient execution attacks on the \ariane \riscv CPU can be prevented with the addition of a single instruction.

\subsection{Attacks on Physical Memory}
Both the \ac{os} and a physical attacker can attempt to access enclave data stored in physical memory.
While a physical attacker could read out enclave  memory via bus probing attacks or cold-boot attacks~\cite{DBLP:conf/IEEEares/LindenlaufHS15}, the \ac{os} has direct access to physical memory.
However, the \ac{os} cannot provide all the necessary tweak information. 
Hence, the attack is mitigated by \rvas detecting the integrity corruption.

Finally, an attacker could install a tampered DRAM module that duplicates the memory on each address and allows to toggle between the two.
This results in a violation of the invariant \IT, and allows for roll-back attacks.
Similarly, the \ac{os} could move around encrypted enclave data and their authentication codes in order to replay stale data.
However, the integrity counters of \rvas protect against roll-back attacks.

\subsection{Attacks on Virtual Memory}

\subsubsection{Memory Isolation.}
Enclaves run in the virtual memory of a host application. 
Thus, the host application or the \ac{os} could try to access enclave data via its virtual address space.
\ourEnclave mitigates these attacks by supplying the memory color field of \rvas with data that is either unforgeable by the \ac{os} (\eg \Mrange, \msidreg{}) or secret, \ie only known by the enclave.
Without a correct tweak, \rvas fails and traps to the \ourSM.

\subsubsection{Page Mapping Attacks.}
The \ac{os} has full control over the \acfp{pte} and can arbitrarily map pages and page permissions. 
Noted as the memory mapping challenge \ChalAddressTranslation (\cf \Cref{sec:security_properties}), this allows for a range of attacks.

\subsubsection{Downgrade Attack.}
A compromised \ac{os} can map an unprotected page to an address in the \Mrange of an enclave and trick it into writing secrets to this unprotected rogue page.
However, when entering the enclave, the \ourSM sets its respective \Mrange registers. 
This register changes the address and segment information supplied to \rvas.
Additionally, the rogue page has not been initialized by the \ourSM.
Thus, the integrity of the page is violated, and the decryption fails.

\subsubsection{Page Remapping Attacks.}
The \ac{os} can attempt to violate invariants \IS and \IT in various ways.
It could remap enclave pages to a different enclave or change the mapping order within an enclave's range.
This remapping allows to manipulate the control flow or divert the data access to different parts of the enclave, \ie to set an encryption key to zero.
Further, since the \acp{pte} also hold the page's permission bits, the \ac{os} could make a data page executable and, for example, exploit a vulnerability in the enclave itself to divert the control flow.

To counteract these attacks, the \rvas tweak includes all relevant mapping information, including a session identifier (\eg the \acf{rtid} or the \acf{encid}), the virtual page offset within an enclave as well as the page permission bits specified by the enclave developer. 
Thus, any deviation from the intended memory mapping results in a decryption error and denies the access.

\subsubsection{Swapping Attacks.}
The \ac{os} can attempt to swap out enclave pages while maintaining the original page intact. 
Without the updated \ac{pte}, the \ac{mmu} will not raise a page-fault during access by the enclave.
Thus, the \ac{os} would have two copies of the same virtual enclave page, which clearly violates our \textit{temporal invariant} \IT. 
\ourEnclave prevents this as follows: the \ourSM overwrites and, thus, invalidates the original page before the swapped-out copy is released to the \ac{os}.

The \ac{os} could try to misuse the swapping mechanism to violate \IT. 
When the attacker requests a page to be swapped, the \ourSM invalidates the physical page and hands over an encrypted copy to the \ac{os}. 
Now, the \ac{os} swaps the page in but keeps the old copy of the swapped page on disk.
The attacker then requests to swap out the same page again. 
Upon the next swap-in operation, the \ac{os} replays the first copy to perform a roll-back attack.
To counteract this attack, the swapped pages are protected with an AEAD scheme.
The authentication tag is linked to the page's virtual address and securely stored on a per-enclave \ptmonitor page.
To enforce \IT, the tag is checked during swap-in, preventing any replays.

\subsubsection{Shared Data Page Attacks.}
Enclave shared memory opens a new attack vector, where the \ac{os} could replace an arbitrary enclave page with a shared memory page (using the zero key), thus tricking the enclave into leaking its secrets.
We close this attack as follows: Accessing shared memory is only possible within the URANGE register, which is initially disabled.
Thus, the enclave has to explicitly configure the URANGE register (after configuring the shared memory key).

The security of shared memory further depends on the \SI{80}{\bit} shared key stored in the \texttt{USID\{0,1\}} \acp{csr}.
A malicious \ac{os} could start a malicious enclave and map the pages of an existing shared memory range to its virtual memory.
This enclave then sets its range registers and brute-forces the shared key as part of an online attack.
Note that this attack can only target shared memory, as the \Xrange separates other enclave page types.

Our \ourSM can prevent brute-force attacks: 
(1) The \ourSM can terminate the enclave that used the wrong shared key after one or a few access attempts.
As spawning a new attacker enclave takes time, this acts as dynamic rate-limiting. 
(2) The \ourSM can perform explicit rate-limiting in its exception handler, thus further reducing the speed of the brute-force attack.

\subsubsection{Shared Code Page Attack.}
\ourEnclave allows different instances of the same enclave to share non-writable pages to deduplicate code and minimize memory usage.
This deduplication is achieved using the \acf{encid} in the tweak as a memory color.
An attacker might try to generate an enclave which yields the same \encid as the victim enclave.
This challenge refers to finding a second pre-image to a cryptographic authentication code.
In a first attack scenario, the attacker could generate a large number of enclaves in an offline brute-force search until the \encid collides with the victim enclave.
Full cryptographic strength (\eg \SI{128}{\bit} security) for the \encid, prevents this attack.
However, \ourEnclave only supports a \SI{80}{\bit} \ac{sid} inside the tweak, which requires truncating the full \encid. 
A simple truncation would drastically facilitate this offline attack.
To counteract this, our \sm performs a key derivation on the \encid that involves the secret CPU key, before truncating it.
In a second scenario, the attacker performs an online brute-force attack by mapping the shared code pages of a victim enclave into the attacker's enclave address space at the same virtual offset. 
If the truncated \acp{encid} match, the attacker enclave can access the shared code pages.
If not, an \authenticationExceptionAsText is raised, and the attacker can retry with a new enclave yielding a different \encid.
As before, the \ourSM can terminate the attacking enclave and apply a time penalty for each \authenticationExceptionAsText.

\section{Evaluation}\label{sec:benchmarks}
Our prototype is based on the \ariane~\cite{DBLP:journals/tvlsi/ZarubaB19} platform consisting of a 64-bit \riscv CPU. 
For \ourEnclave, we extended this platform with the \rvas ISA extensions, the storage of tweaks in the cache, and a \ac{mee} for \rvas.
The \ac{isa} extensions of \ourEnclave allow the security monitor to set the tweak in software.
Here, we endow \ariane with additional CSRs and a tweak logic, as depicted in \Cref{fig:rvas}.
We further extend the write-through cache to handle the tweak, \ie storing these bits next to the data entries and considering the tweak in the hit logic.
We increased the default cache line size from \SIrange{16}{64}{\byte}, a common choice for many CPUs.
We use \memsec~\cite{DBLP:conf/fpl/WernerUSSM17}, an open-source, AXI compatible framework supporting various encryption schemes for the \ac{mee}.
To fulfill our requirements (\cf ~\Cref{sec:rvas}), we configured \memsec to use ASCON-128.
We use this cryptographic primitive for \rvas because it is the only cipher that is supported by the \memsec framework in TEC-Tree mode. 
Furthermore, we extend \memsec to process the tweak as \ascon's associated data.
\memsec is placed between the cache and the memory controller to transparently encrypt all data leaving the processor.
We transport the tweak from the core to \memsec using the user-defined signals of the AXI4 communication fabric. 

\subsection{Performance Overhead}
\begin{figure}
	\centering
	\begin{subfigure}[t]{0.48\textwidth}
		\includegraphics[width=\textwidth]{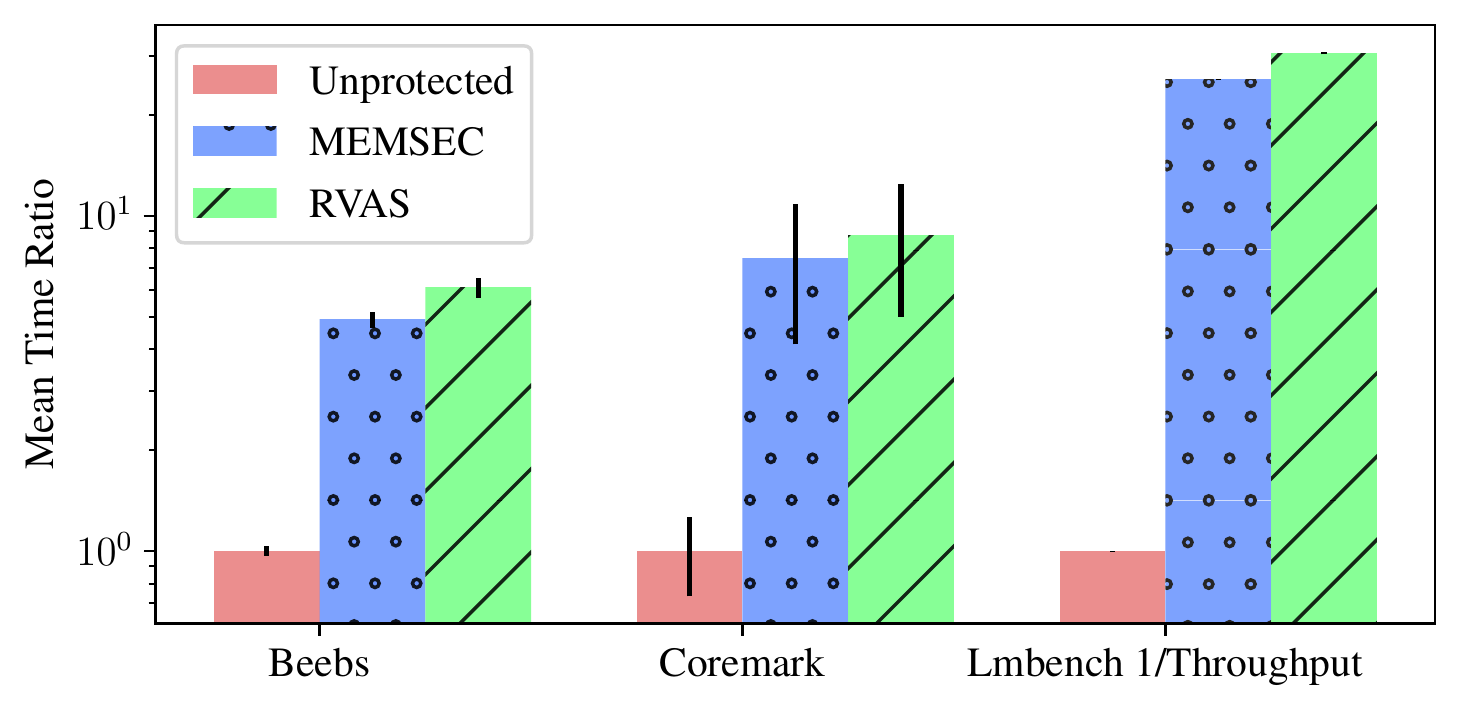}
		\vspace{-0.5em}
		\centering
		\captionof{figure}{Benchmark results for the \ariane core using no memory encryption, \memsec, or \rvas. Numbers in red are given as ratios to the unprotected configuration.}
		\label{fig:benchmarkresult}
	\end{subfigure}
\end{figure}

To measure the performance overhead introduced by \rvas, we deployed it on a Xilinx Kintex-7 FPGA KC705 board and ran a variety of macrobenchmarks on a Linux 5.10 kernel. 
To simulate different workloads, we use BEEBS~\cite{DBLP:journals/corr/PallisterHB13} and CoreMark~\cite{coremark}.
For BEEBS, we excluded the \texttt{crc32}, \texttt{ludcmp}, \texttt{st}, \texttt{matmult-float}, and \texttt{rijndael} benchmarks, since they caused lockups on the unmodified \ariane CPU.
For the results in \Cref{fig:benchmarkresult}, we aggregated all BEEBS benchmarks into a single metric by using the geometric mean, while the full BEEBS results are given in the \Cref{app:beebs_complete} in \Cref{fig:beebs_complete}.
For the fast-running CoreMark, we plot the mean over \numprint{1000} runs, while for the slower BEEBS, we average over 25 runs.
For each run, CoreMark uses 10 internal iterations, while for BEEBS, we use 4 internal iterations.
Due to the resource-constrained prototype (\SI{256}{\mega\byte} accessible DRAM and \SI{50}{\mega\hertz} CPU frequency), we cannot run more heavyweight benchmarks.

\begin{figure}
	\begin{subfigure}[b]{0.48\textwidth}
		\includegraphics[width=\textwidth]{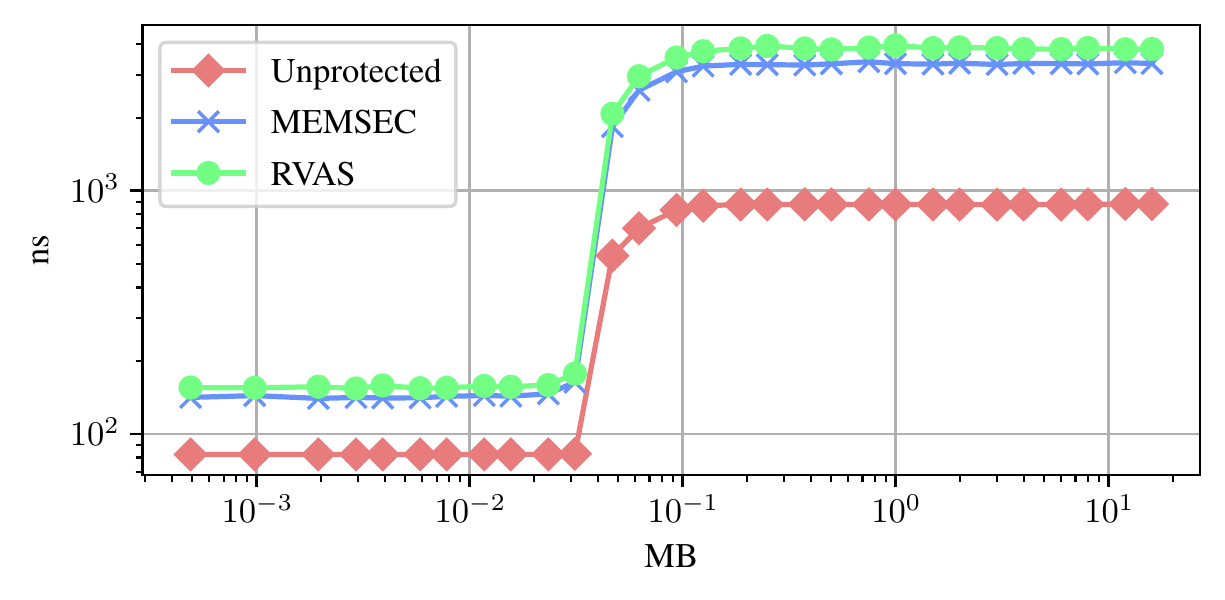}
		\vspace*{-0.5cm}
		\centering
		\captionof{figure}{Memory read-write latency.}
		\vspace{-0.5em}
		\label{fig:memlatency}
	\end{subfigure}
	\captionof{figure}{Benchmarking and memory latency results on the \ariane platform in an unprotected, \memsec and \rvas configuration.}
\end{figure}

\Cref{fig:benchmarkresult} depicts the results of our evaluation normalized to an unprotected baseline, \ie the \ariane core without any memory encryption.
The use of stock \memsec configured for an authentication tree constitutes the main overhead for all benchmarks.
As we will discuss in \Cref{sec:limitations}, the memory encryption overhead can be significantly reduced in practice.
The overhead of \rvas over \memsec is caused by the additional rounds of \ascon we use to process the tweak as part of its associated data. Here, \rvas adds two additional calls to \ascon's permutation function in the \ac{mee}.
The \rvas prototype adds an overhead between \SI{\macrooverheadcoremark}{\percent} for CoreMark, \SI{\macrooverheadthroughput}{\percent} for LMbench~\cite{DBLP:conf/usenix/McVoyS96} and \SI{\macrooverheadbeebs}{\percent} for BEEBS compared to \memsec.
\Cref{fig:memlatency} shows the results of the read-write latency test of LMbench for different configurations.
More concretely, this benchmark measures the read-write latency for different sized data chunks and visualizes the impact of the \SI{32}{\kilo\byte} L1 data cache of \ariane and the latency of the external DDR3 memory.
While \memsec increases the average read-write latency for a memory access from \SIrange{850}{3300}{\nano\second}, the two additional rounds of \rvas only increase the latency by additional \SI{290}{\nano\second} on average.
These results are encouraging, given that we instantiated our \rvas prototype with the general-purpose \memsec encryption framework.
An encryption engine that is tailored towards \rvas (\eg by optimizing block sizes) will further reduce the overhead.
We discuss possible optimizations in \Cref{sec:limitations}.

\begin{table}[h]
	\setlength{\aboverulesep}{0pt}
	\setlength{\belowrulesep}{0pt}
	\caption{Micro-benchmarking results for \ourEnclave.}
	\label{tab:benchmarks}
	\centering
		\resizebox{1\columnwidth}{!}{%
	\begin{tabu}{lrr}
		& cycles median & relative to getpid\\
		\midrule
		Syscall getpid                     & \numprint{\benchriscvgetpid}    & \benchriscvgetpidgetpid \\
		\ourEnclave SM Call \enquote{null} & \numprint{\benchriscvsmnull}    & \benchriscvsmnullgetpid \\
		\ourEnclave Enter                  & \numprint{\benchriscventer}     & \benchriscventergetpid \\
		\ourEnclave Exit                   & \numprint{\benchriscvexit}      & \benchriscvexitgetpid \\
		\ourEnclave Create                 & \numprint{\benchriscvcreate}    & \benchriscvcreategetpid \\
		Context Switch Sem.                & \numprint{\benchriscvsemaphore} & \benchriscvsemaphoregetpid \\
	\end{tabu}
		} % resizebox end
\end{table}

Furthermore, we evaluate \ourEnclave using the microbenchmarks shown in \Cref{tab:benchmarks} on \ariane equipped with \rvas.
To reduce the scheduling- and cache-related differences in the measurement results, we repeat each test \numprint{10000} times.
To get a sense of the switching overhead of \ourEnclave, we measure the number of cycles an \eenter/\eexit takes and compare it with a simple \enquote{getpid} system call.
The \enquote{null SM call} is the equivalent of a getpid system call, but instead of calling into the \ac{os}, we invoke the \sm.
Any \sm-related action (e.g., \eenter/\eexit) also includes this baseline overhead for an SM call.
Calling an enclave function only takes \benchriscventerexitgetpid{} the time of a simple system call.
This call includes the time for entering (\numprint{\benchriscventer} cycles), executing a simple function within the enclave which only returns a fixed value, as well as exiting (\numprint{\benchriscvexit} cycles) the enclave.
We implemented a context-switch benchmark using a semaphore and shared memory for synchronization to compare this with traditional process-based isolation. 
Process-based context switching takes \benchriscvsemaphoregetpid{} of a simple system call.
For comparison, entering and exiting an Intel SGX enclave takes 71x the time of a system call~\cite{DBLP:conf/eurosys/KoningCBGA17}, thus being twenty times slower than invoking a \ourEnclave enclave.
This result highlights the benefit of \ourEnclave not requiring an expensive \ac{tlb} flush~\cite{IntelSDM,DBLP:journals/iacr/CostanD16} when invoking an enclave.

\subsection{Hardware Overhead}\label{sec:hwoverhead}
The hardware overhead of \rvas consists of the \ac{mee}, the ISA extension, and the cache with the additional tweak bits.
We synthesize our modified \ariane for a Xilinx Kintex-7 series FPGA.
Compared to the default \ariane, the design increases by \SI{\hwoverheadoveralllut}{\percent} in terms of lookup tables (LUTs) and \SI{\hwoverheadoverallff}{\percent} in terms of flip-flops.
From these \SI{\hwoverheadoveralllut}{\percent} additional LUTS, \SI{\hwoverheadoveralllutmee}{\percent} result from the introduced \ac{mee}, \SI{\hwoverheadoveralllutcache}{\percent} of the extended cache, and the rest by the ISA extension.
\rvas requires that each cache line is tagged with \SI{125}{\bit} for the memory encryption tweak due to \ariane implementing \SI{39}{\bit} virtual address spaces. 
While the tag overhead depends on the design of the cache and the size of the cache lines, for our 512-bit cache lines, the overhead is \SI{25}{\percent}.
However, this overhead could be reduced using the optimizations in \Cref{sec:tweakCache}.

\subsection{Prototype Limitations}\label{sec:limitations}
Our \rvas design, as a prototype implementation, is not optimized for performance.
Due to a lack of openly available high-performance \ac{mee} that support authentication, we used the \memsec~\cite{DBLP:conf/fpl/WernerUSSM17} framework. 
As seen in \Cref{fig:benchmarkresult}, the \ac{mee} is significantly responsible for the overall performance overhead.
According to ARM, full memory encryption induces a runtime overhead of \SIrange{7.5}{25}{\percent}  and a storage overhead of \SIrange{7.8}{26.7}{\percent}~\cite{ARMMemEncPresentation}. 
Different workloads on Intel SGX may run up to 3--19x slower~\cite{DBLP:conf/isca/MofradZLS18}.
Given recent advances in the \riscv community, we also expect high-performance \acp{mee} to become openly available in the future.
Currently, our prototype transparently encrypts the whole external DRAM.
An encryption bypass could selectively disable encryption for unprotected data and improve the system performance. 

\subsubsection{Caches.}
The \ariane platform we used for the evaluation features a write-through cache that slows down write operations on encrypted memory.
Adopting a write-back cache could significantly improve write performance.
In our current prototype implementation, every cache line is tagged with the full tweak, yielding the area overhead mentioned in \Cref{sec:limitations}.
To reduce this storage overhead to a minimum, a dedicated \acl{tc}, as elaborated in \Cref{sec:tweakCache}, could be installed. 

\section{Related Work}\label{sec:related_work}
\subsubsection{\intel SGX~\cite{DBLP:conf/isca/McKeenABRSSS13,DBLP:journals/iacr/CostanD16}}
is a set of instructions to manage and interact with enclaves on the x86 platform.
While SGX enclaves use the same userspace virtual memory as the accompanying application, the pages, which can be up to \SI{128}{\mega\byte} and are only accessible by the enclave, are located in the \textit{Processor Reserved Memory} (PRM) region in the memory.
The PRM holds the \SI{4}{\kilo\byte} pages belonging to the enclave in the \textit{Enclave Page Cache}, and the \textit{Enclave Page Cache Map} (EPCM) is a trusted metadata storage storing the state of the EPC.
In contrast to SGX, \ourEnclave is not limited to a statically allocated memory region (\ie the PRM) but dynamically reuses the whole physical memory for both unprotected code and enclaves. 
Furthermore, our approach does not require a trusted metadata storage such as the EPCM but instead feeds this metadata directly into the encryption.
Finally, \ourEnclave does not require flushing of the \ac{tlb}.

\subsubsection{CrypTag~\cite{DBLP:journals/corr/abs-2012-06761}}
assures safety for memory allocations by utilizing the unused upper bits of pointers to supply a tag to a memory encryption for specific instruction or data accesses.
In contrast, RVAS supports various policies and incorporates information on the CPU state or specified by a \ourSM (\cf \Cref{sec:rvas}).
On the hardware side, RVAS adds all the necessary logic to enforce these policies. 
The cache area overhead of CrypTag is up to \SI{20}{\percent}, which is comparable with RVAS.

\subsubsection{VAULT~\cite{DBLP:conf/asplos/TaassoriSB18}}
aims to reduce the paging overhead by making the EPC of \intel SGX available to the full system memory.
Unlike \ourEnclave, VAULT does not overcome the limitation of SGX regarding efficient shared memory.

\subsubsection{SMARTS~\cite{DBLP:conf/isca/WongHC18}}
implements a \acf{mpu} as a framework that can perform partial memory encryption. 
The physical DRAM is partitioned into an untrusted, a trusted, and a metadata region.
In contrast to SMARTS, \ourEnclave is not bound to a static boot-time memory configuration and allows fully dynamic management of enclave pages at runtime.

\subsubsection{\amd Secure Encrypted Virtualization (SEV)~\cite{AMDSev,AMDMemoryEncryption,AMDSevEs,AMDSevSnp}}
is a set of CPU extensions to execute virtual machines in untrusted environments.
\amd SEV~\cite{AMDSev} comprises secure memory encryption (SME)~\cite{AMDMemoryEncryption}, SEV-encrypted state (SEV-ES)~\cite{AMDSevEs}, and SEV-secure nested paging (SEV-SNP)~\cite{AMDSevSnp}.
Combined, they allow for protection against memory replay, data corruption, memory aliasing, and memory re-mapping attacks.
In contrast to AMD-SEV, \ourEnclave focuses on enclaves rather than virtual machines.
Moreover, \amd-SEV does not offer integrity protection against physical attacks.

\subsubsection{Intel MKTME~\cite{IntelMKTME}}
is a proposal to transparently encrypt memory pages.
Based on the \ac{pte}, one out of 64 different encryption keys can be selected.
Similar to SEV, MKTME does not provide cryptographic authentication.
Unlike SEV, it needs to rely on a trusted hypervisor.

\subsubsection{Other systems.} 
Sanctum~\cite{DBLP:conf/uss/CostanLD16}, Keystone~\cite{DBLP:conf/eurosys/LeeKSAS20}, CURE~\cite{bahmani2020cure}, and TrustZone~\cite{ARMTrustZone} are other recent enclave and TEE designs tackling challenges such as cache-based attacks and enclave-to-peripheral binding.
However, in contrast to \ourEnclave, all these designs do not explicitly protect the external memory from physical attacks using memory encryption.

\section{Future Work}\label{sec:additionalFeatures}
We see usage scenarios of \rvas beyond traditional enclaves to provide, for example, fine-grained intra-enclave isolation, and system-level enclaves.
\ourEnclave could be used to supersede other protection mechanisms such as memory protection keys~\cite{IntelSDM}, pointer authentication~\cite{DBLP:conf/uss/LiljestrandNWPE19}, pointer tagging~\cite{ARMMTEWhitepaper}, and memory coloring~\cite{DBLP:journals/corr/abs-2012-06761}.
Furthermore, \rvas presents a building block that could be used to enable secure virtualization (\cf \amd-SEV) without the need for different encryption keys and with additional protection against physical attacks.
\ourEnclave specifies a number of configuration registers on each privilege level.
These registers can allow for additional protection in the kernel by creating kernel-level enclaves.
Our current prototype implementation uses ASCON as it is a lightweight cryptographic primitive already available in \memsec.
However, realizing \rvas with other encryption primitives, such as AES, would be possible but requires additional analysis, which we leave open for future work.

\subsubsection{Protection against malicious enclaves.}
While outside the threat model of enclaves, the host application can be protected from a potentially malicious enclave~\cite{DBLP:conf/dimva/0001WG19} using techniques similar to SGXJail~\cite{DBLP:conf/raid/WeiserMSG19}.
\rvas' memory colors could be leveraged for this purpose. 

\section{Conclusion}\label{sec:conclusion}

This paper presented an innovative isolation primitive called authenticryption shield that unifies traditional and advanced isolation policies and offers potential for future security applications.
We illustrated how it streamlines security reasoning by building on top of a tweakable memory encryption scheme, thus giving cryptographic isolation guarantees.
We demonstrated how the versatility of our approach allowed us to design and prototype an innovative and novel enclave architecture for \riscv called \ourEnclave, that even also allows for native and secure sharing between enclaves. 
As a generic extension for the \riscv \ac{isa}  and a small Security Monitor, we showed how \ourEnclave, at a size of just \locsmcomplete~\acp{loc}, can manage all enclaves throughout their life-cycle.
Additionally, we thoroughly assessed \ourEnclave's performance and showed that entering or exiting takes only about \benchriscventerexitgetpid{} of a getpid syscall.
We prototyped and evaluated \ourEnclave on the \ariane RISC-V hardware and plan to make the prototype publicly available.

\ifanonymous
\else
%-------------------------------------------------------------------------------
\section*{Acknowledgments}
%-------------------------------------------------------------------------------
This project has received funding from the European Research Council (ERC) under the European Union’s Horizon 2020 research and innovation programme (grant agreement No 681402) and by the Austrian Research Promotion Agency (FFG) via the competence center Know-Center (grant number 844595), which is funded in the context of COMET - Competence Centers for Excellent Technologies by BMVIT, BMWFW, and Styria. Furthermore, this work has been supported by the Austrian Research Promotion Agency (FFG) via the project ESPRESSO, which is funded by the province of Styria and the Business Promotion Agencies of Styria and Carinthia.
\fi % ifanonymous

\FloatBarrier
\bibliographystyle{plain}
\bibliography{./bibliography}

\clearpage

\appendix\label{sec:appendix}

\section{\ourSM API}\label{app:sm_api}

The \sm provides a number of API functions that allow applications, the \ac{os}, or the enclave to interact with the \sm.
Therefore, we define the following API functions offered by our \sm:

\subsubsection{\ecreate} is used for loading and initializing an enclave. 
It is called with the virtual location of the encrypted enclave code, its virtual target location, a memory area for the enclave stack, and two pages used for storing per-enclave and per-thread metadata. 
Once invoked, the \sm decrypts the enclave and initializes its stack and metadata pages, using the tweak override. A hash of the decrypted enclave is stored in the per-enclave page for later use (\eg \egetsealkey).

\subsubsection{\eenter} is called by the user-mode application with the per-enclave and per-thread pages generated by \ecreate.
When calling \eenter, the \sm saves relevant CPU registers on the per-enclave page,
initializes the enclave \acp{csr}, and starts the enclave code at its predefined entry-point.

\subsubsection{\eexit} is called by the active enclave to return to the user application.
Here, the \sm restores the register state from before entering the enclave (apart from return value registers) and resets the enclave's \acp{csr}.
Finally, the program counter is set such that the program continues after its last \eenter call.

\subsubsection{Interruption.}
When an enclave is interrupted, the \sm stores the enclave register state in its per-thread page, wipes the registers, and hands over execution to the \ac{os}.
When the interrupted process invokes \eenter again, the \sm restores the saved enclave state and resumes execution. 

\subsubsection{\eprepare} makes the \sm initialize and zeroes an enclave page to an enclave-defined page type using the tweak override.
It allows for any page type apart from the \ptmonitor.
This allows an enclave to initialize pages dynamically allocated by the user-mode application.

\subsubsection{\edestroy} makes the \sm destroy a physical enclave page by choosing an otherwise unused value for the tweak override before writing the entire page.

\subsubsection{\emod} offers in-place re-encryption of a page. 
Called with the virtual address of the page, the old page type, and the new page type. 
This call can be used to deliberately change \pt permission bits or to rekey a shared \ptshm page.
This call has to be made if an mprotect or mmap syscall altered tweak-relevant bits of the PTE.
The \sm closely follows the procedure from \eprepare, but preserves the pages content.

\subsubsection{\egetsealkey} is used to acquire a deterministic sealing key derived from the enclave hash and the CPU key.
This call allows the enclave to safely encrypt its secrets for persistent storage. 

\section{Cache Eviction Probability}\label{app:eviction_probability}
\begin{figure*}
	\centering
	\begin{subfigure}[t]{0.34\textwidth}
		\centering
		\includegraphics[width=\textwidth]{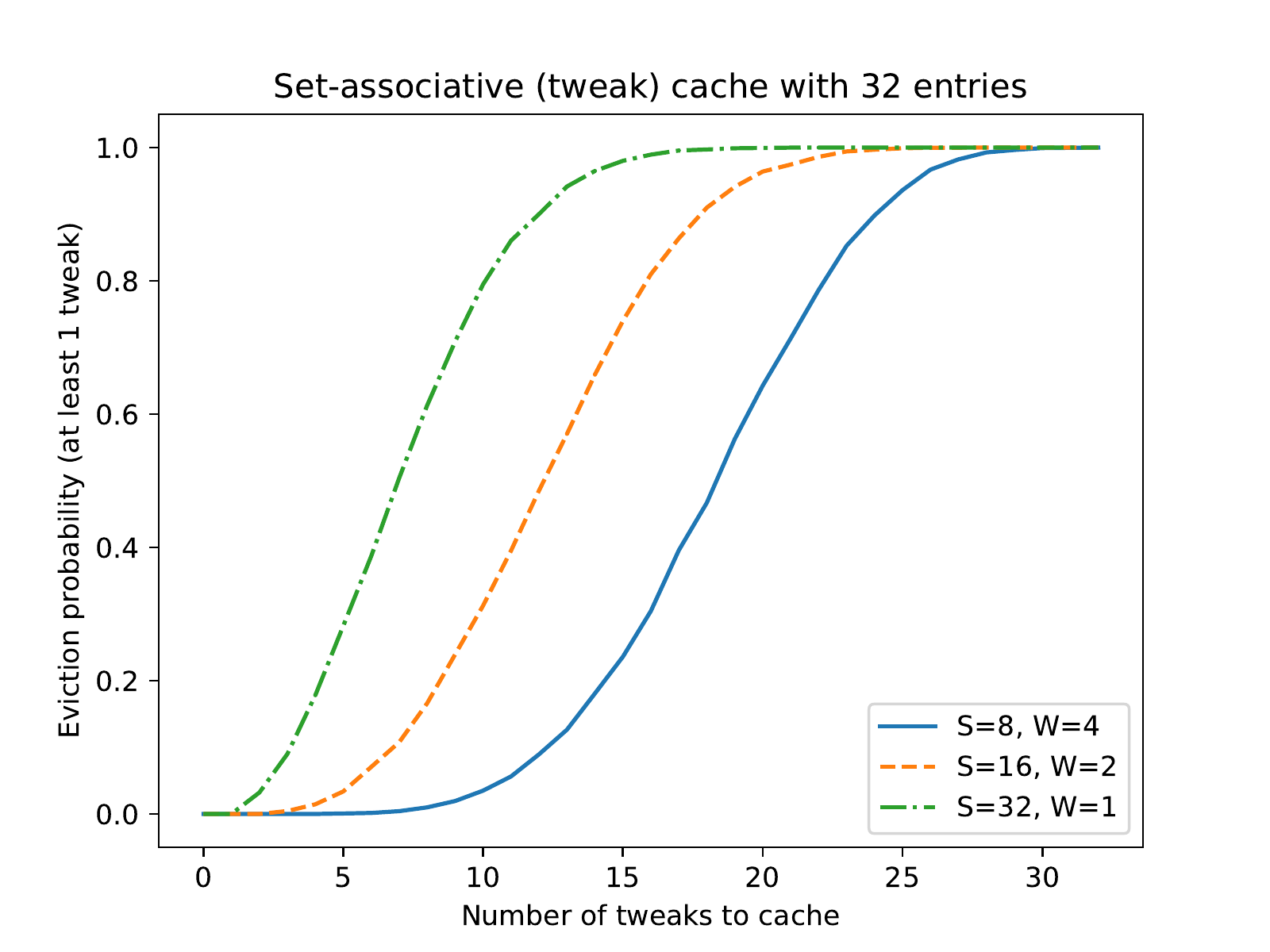}
		\captionof{figure}{\ac{tc} with 32 entries.}
		\label{fig:tweakevictionprobability32}
	\end{subfigure}
	\vspace{-1.5em}
	\begin{subfigure}[t]{0.34\textwidth}
		\centering
		\includegraphics[width=\textwidth]{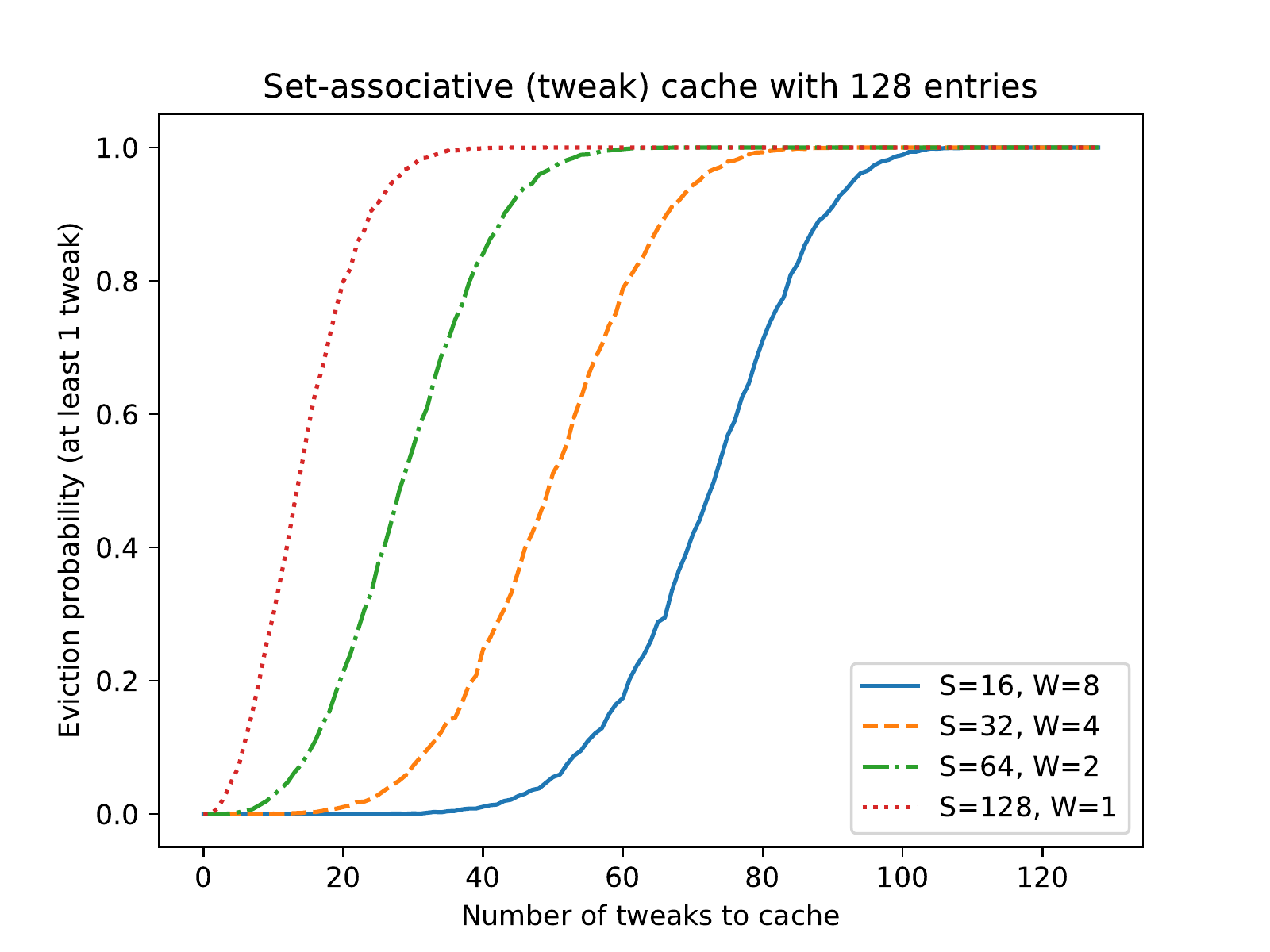}
		\captionof{figure}{\ac{tc} with 128 entries.}
		\label{fig:tweakevictionprobability128}
	\end{subfigure}
	\captionof{figure}{The probability of evicting at least one entry for tweak caches with 32 or 128 entries and 1, 2, 4  or 8 ways.}
	\label{fig:tweakevictionprobability}
\end{figure*}

\begin{figure*}
	\centering
	\begin{subfigure}[b]{0.34\textwidth}
		\centering
		\includegraphics[width=\textwidth]{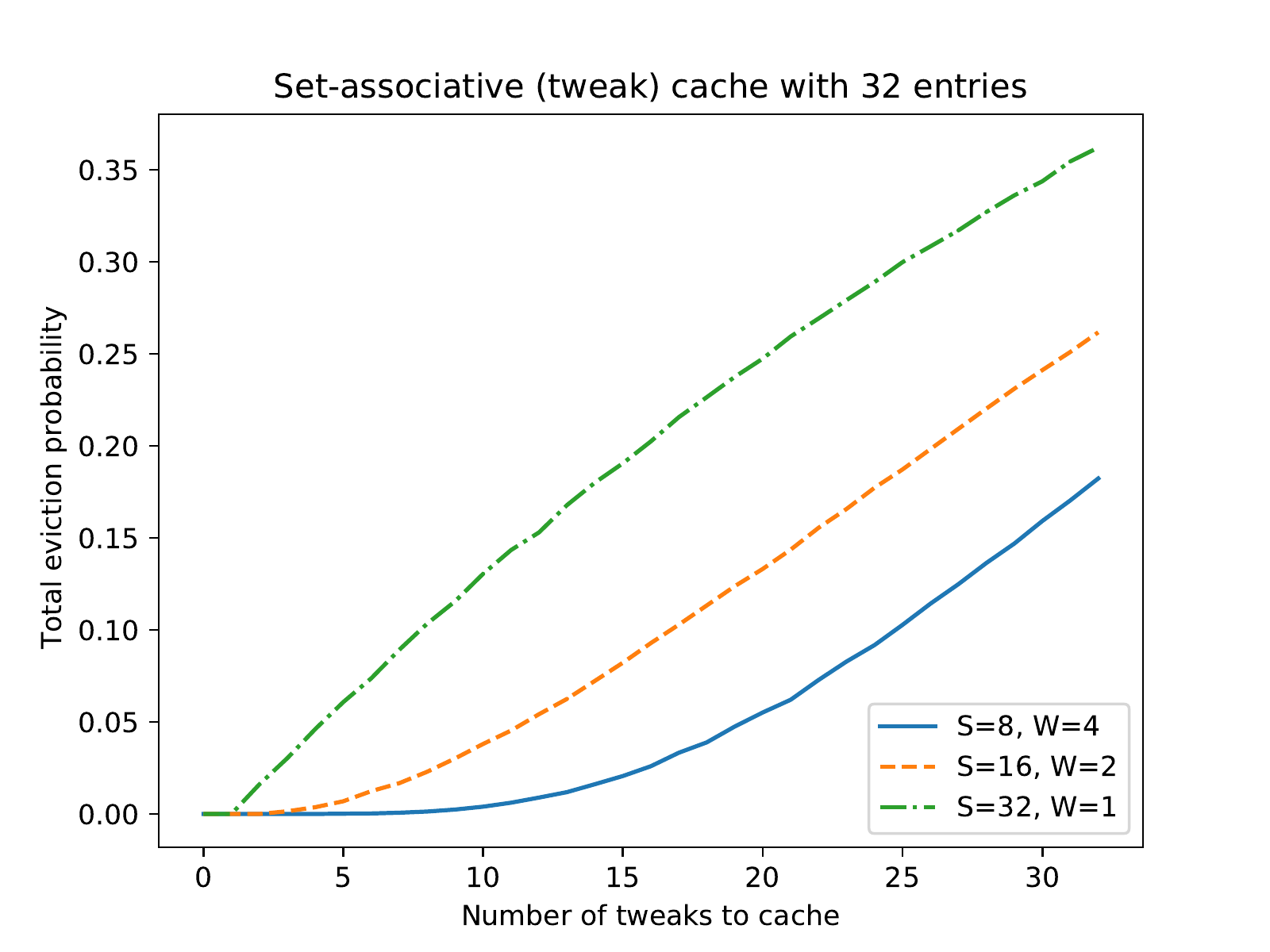}
		\captionof{figure}{\ac{tc} with 32 entries.}
		\label{fig:totaltweakevictionprobability32}
	\end{subfigure}
	\vspace{-1.5em}
	\begin{subfigure}[b]{0.34\textwidth}
		\centering
		\includegraphics[width=\textwidth]{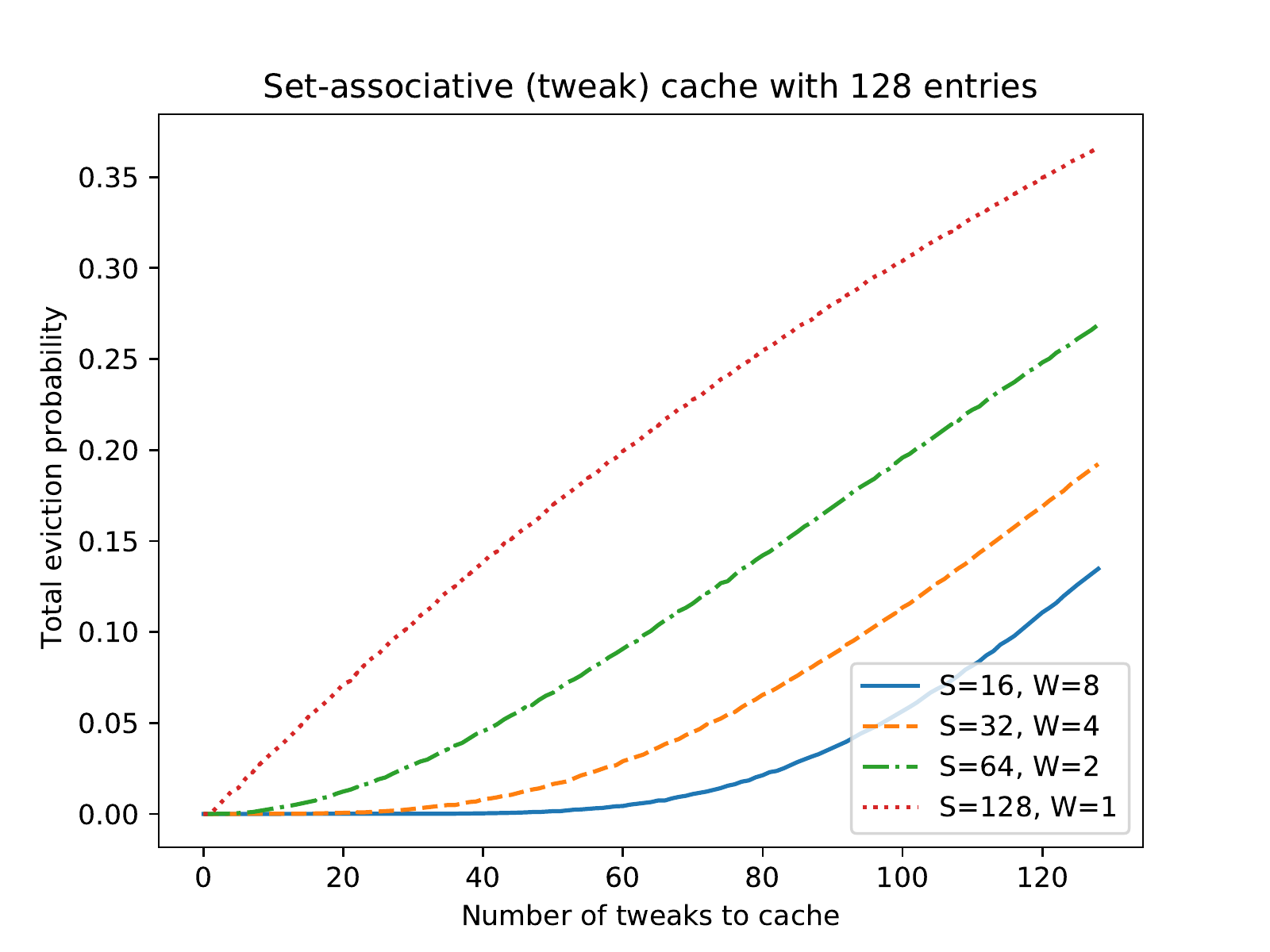}
		\captionof{figure}{\ac{tc} with 128 entries.}
		\label{fig:totaltweakevictionprobability128}
	\end{subfigure}
	\captionof{figure}{The total eviction probability for tweak caches with 32 or 128 entries and 1, 2, 4  or 8 ways.}
	\label{fig:totaltweakevictionprobability}
\end{figure*}

Intuitively, the size of a \acf{tc} and its number of ways determine its eviction probability. 
This probability corresponds to the likelihood that, \eg one enclave experiences self-eviction.
To do so, we simulate the eviction probability with two experiments as follows:
First, we simulate the probability that at least one tweak is evicted when accessing a certain number of tweaks in \Cref{fig:tweakevictionprobability}.
Second, we simulate the total eviction probability, \ie the probability that a tweak is evicted when accessing a certain number of tweaks in \Cref{fig:totaltweakevictionprobability}.
Both experiments are evaluated with random set indices for our \ac{tc}, which approximates the distribution of the cryptographic index derivation function.
We use \acp{tc} with capacities of 32 and 128 entries and one to eight ways.
We repeat each experiment 10000 times.
As seen in \Cref{fig:tweakevictionprobability} and \Cref{fig:totaltweakevictionprobability}, the eviction probability of a tweak entry can be significantly reduced by increasing the number of ways or increasing the number of possible tweak entries.

To put this into perspective, we assume that one enclave uses 6 tweaks, and we can accept a total eviction probability for our system of only \SI{5}{\percent}.
With parameters in mind, we can run roughly 2 enclaves in parallel on a system with a small \ac{tc}, \ie with only 32 entries and 2 ways, with low probability of the enclaves interfering with each other, as seen in \Cref{fig:totaltweakevictionprobability32}.
On the larger configuration in \Cref{fig:totaltweakevictionprobability128} with 128 entries and 4 ways, about 11 enclaves could run with the same low eviction probability.

\section{Detailed Evaluation Results}\label{app:beebs_complete}
\begin{figure}[h]
	\includegraphics[width=0.90\textwidth]{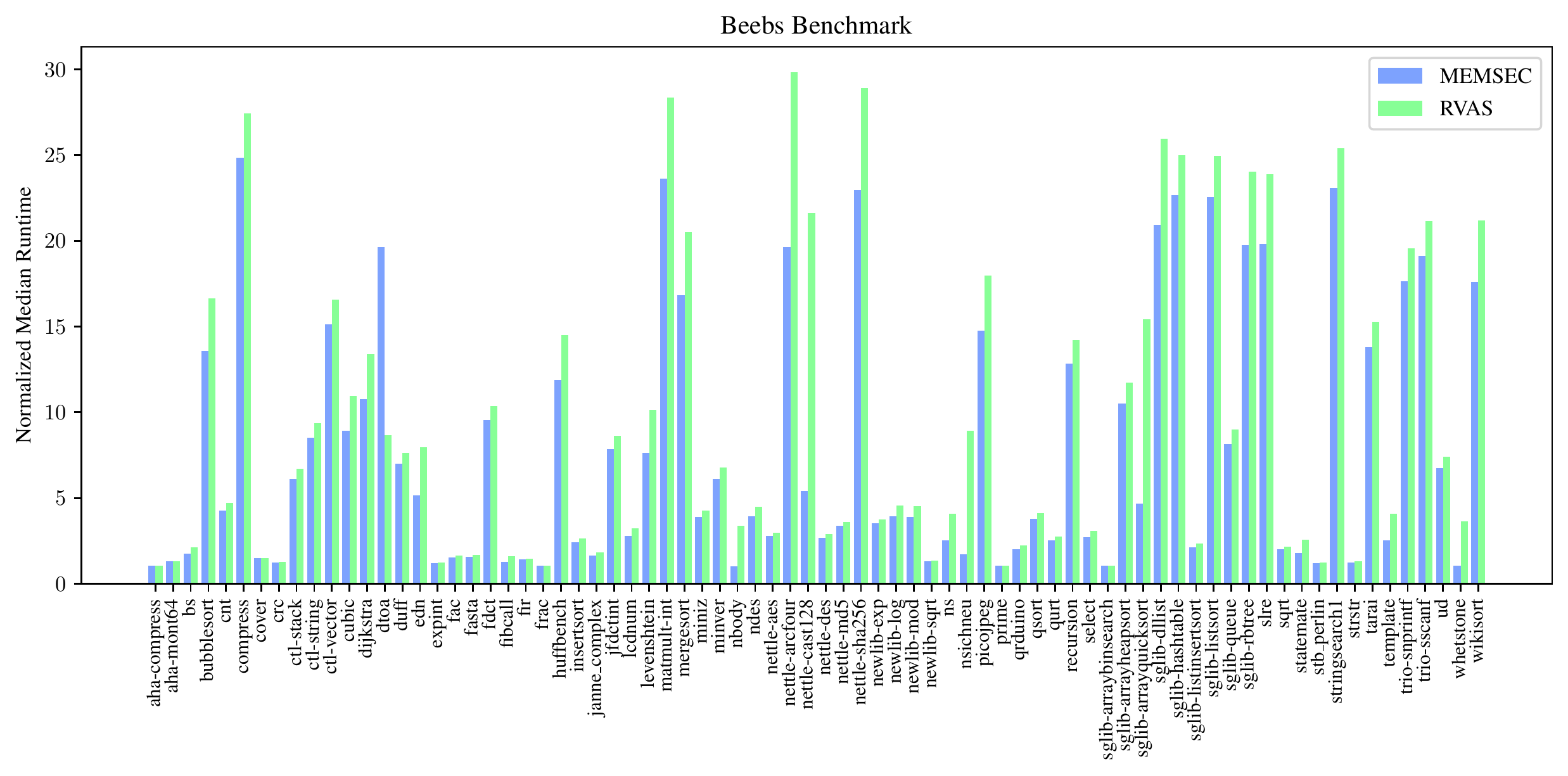}
	\captionof{figure}{Individual results of the BEEBS benchmark suite, normalized to an unprotected implementation.}
	\label{fig:beebs_complete}
\end{figure}

\Cref{fig:beebs_complete} shows the individual results of the BEEBS benchmark suite when run on \rvas, compared to the MEMSEC encryption engine we used.
Both are normalized relative to an unprotected implementation (\ie without enclaves or memory encryption).

\FloatBarrier

%%%%%%%%%%%%%%%%%%%%%%%%%%%%%%%%%%%%%%%%%%%%%%%%%%%%%%%%%%%%%%%%%%%%%%%%%%%%%%%%
\end{document}